\documentclass{IEEEtran}
\usepackage{tikz}

\usetikzlibrary{external}
\usepackage{cite}
\usepackage{multicol}
\usepackage{amsmath,amssymb,amsfonts}
\newcommand\ddfrac[2]{\frac{\displaystyle #1}{\displaystyle #2}}
\usepackage{graphicx}
\usepackage{caption,float,color,subcaption}
\usepackage{textcomp}

\usepackage{algorithm,cuted}
\usepackage[noend]{algpseudocode}

\usepackage[space]{grffile} 
\usepackage{chngpage}
\usepackage{calc}

\usepackage{tikz-3dplot}
\usepackage[utf8]{inputenc}
\usepackage{pgfplots} 
\usepackage{pgfgantt}
\usepackage{pdflscape}
\pgfplotsset{compat=newest} 
\pgfplotsset{plot coordinates/math parser=false}
\pgfplotsset{every  tick/.style={black,},ylabel style={font=\tiny},xlabel style={font=\tiny},tick label style={font=\tiny},legend style= {font=\scriptsize},
minor x tick num=1,minor y tick num=1,xminorticks=true,yminorticks=true,}
  \newlength\fheight
\newlength\fwidth

\makeatletter
\def\BState{\State\hskip-\ALG@thistlm}
\makeatother

\makeatother
\def\BibTeX{{\rm B\kern-.05em{\sc i\kern-.025em b}\kern-.08em
    T\kern-.1667em\lower.7ex\hbox{E}\kern-.125emX}}
\begin{document}
\title{Sub-6 GHz Microstrip Antenna: Design and Radiation Modeling}
\author{Elyes Balti, \IEEEmembership{Student Member, IEEE}, Brian K. Johnson, \IEEEmembership{Fellow, IEEE}
}

\maketitle

\begin{abstract}
This paper presents a global framework analysis of a microstrip antenna design circularly polarized with an operating frequency of 5.80 GHz and a bandwidth of at least 500 MHz. We evaluate the optimal antenna parameters to design requirements. Capitalizing on these parameters, we simulate the radiation model of this antenna using the Finite Difference Time Domain (FDTD) technique assuming one, two, and three dimensions. The propagation medium is assumed to be a free space bounded by absorbing boundaries, and perfect matched layer (PML). The FDTD-1D is considered in free space while FDTD-2D and 3D are considered both in free space and in a free space-medium containing either dielectric sphere or cylinder in the center. In this case, we model the incident and the scattered electromagnetic fields reflected back from hitting the dielectric object. Moreover, the microstrip antenna radiates an electromagnetic pulse either in the middle or at one end of the medium and the sources considered are Gaussian pulse and plane wave. Finally, we provide the analytic solutions of the propagation models to confirm the accuracy of the FDTD simulation technique.
\end{abstract}



\section{Introduction}
\IEEEPARstart{T}{he} Finite-Difference Time-Domain (FDTD) has become today's one of the most popular techniques for the solution of electromagnetic problems. It has been successfully
applied to an extremely wide variety of problems, such as scattering from metal objects and dielectrics, antennas, microstrip circuits, and electromagnetic absorption in the human body exposed to radiation. The main reason of the success of the FDTD method resides in the fact that the method itself is extremely simple, even for programming a three-dimensional code. The technique was first proposed by K. Yee, and then improved by others in the early 70s \cite{10,11,12,13}. 
\subsection{Microstrip Antenna Overview}
An antenna is an electrical conductor or a system of conductors which is part of a transmitting or receiving system that is designed to radiate or receive electromagnetic waves \cite{1}.\\
A microstrip antenna consists of a thin metallic conductor which is bonded to thin grounded dielectric substrates. The size miniaturization of a microstrip patch antenna is crucial in many of the modern day practical applications, like that of WLAN \cite{2,3}, Wi-Fi \cite{4}, and Bluetooth \cite{5}. Patch antennas play a very significant role in today's world of wireless communication systems. A micro strip patch antenna is relatively simple in construction and makes use of a conventional microstrip fabrication technique which is comprised of the etching of the antenna element pattern in a metal trace which is bonded to an insulating dielectric substrate, such as a printed circuit board (PCB), with a continuous metal layer bonded to the opposite side of the substrate which acts as the ground plane. The most commonly used microstrip patch antennas are rectangular patch antennas, but even circular patch antennas are widely used.\\
Microstrip patch antennas possess a very high antenna quality factor (Q), where a large Q would lead to a narrow bandwidth and low efficiency. The factor Q can be reduced by increasing the thickness of the dielectric substrate but as the thickness increases there will be a simultaneous increase in the fraction of the total power delivered by the source into a surface wave which can be effectively considered as an unwanted power loss since it is ultimately scattered at the dielectric bends and causes degradation of the antenna characteristics. Other problems such as lower gain and lower power handling capacity can be overcome by using different methods. One technique is using an array configuration for the elements which is a collection of homogeneous antennas oriented similarly to get greater directivity and gain in a desired direction. The inset-fed microstrip antenna provides impedance control with a planar feed configuration.
\subsection{Contributions}
In this work, we propose an optimal design of a resonant microstrip antenna design operating at 5.80 GHz following some specifications such as achieving a bandwidth of 500 $BW$ and characterized by a circular polarization.
Capitalizing on this design, we refer to FDTD 1-D, 2-D, and 3-D to describe the radiation model in various media. The analysis of this paper follows these steps:
\begin{enumerate}
    \item Finds the optimum values that satisfy the design requirements.
    \item Presents figures of merit of the antenna such as current distribution, VSWR, impedance, radiation patterm, directivity, and return loss, etc in order to check the accuracy of the design values.
    \item Generates a Gaussian pulse and provides the FDTD formulations in 1D, 2D, and 3D in various media refering to Maxwell's equations.
    \item Provides the analytic solutions to validate the FDTD simulations.
\end{enumerate}
The rest of this paper is organized as follows: Section II
describes the antenna design while the
FDTD-1D analysis is given in Section III. FDTD-2D and 3-D results are detailed in Sections IV, and V, respectively.
Finally, concluding remarks and future research directions are presented in Section VI.
\section{Antenna Design}
In this section, we consider a rectangular patch antenna. The operating frequency of the antenna is  $f_0$ = 5.80 GHz with a bandwidth of at least $BW$ = 500 MHz. The antenna should have a circular polarization or as close to hat as possible. Now, we must choose the relaistic materials for the antenna and substrate.\\
Assuming resonant antenna, the width of the patch is given by \cite{6,7,8}
\begin{align}
W = \frac{c}{2f_0}\sqrt{\frac{2}{\epsilon_{\rm{r}} + 1}},    
\end{align}
where $c$ and $\epsilon_{\rm{r}}$ are the speed of light and the dielectric substrate, respectively.\\
The effective dielectric constant can be given by \cite{6,7,8}
\begin{equation}
\epsilon_{\rm{reff}} = \frac{\epsilon_{\rm{r}} + 1}{2} + \frac{\epsilon_{\rm{r}} - 1}{2}\sqrt{1+12\frac{h}{W}},~~~W > h,
\end{equation}
where $h$ is the height of the substrate.\\
Due to fringing, electrically the size of the antenna is increased by an amount of $\Delta L$. Therefore, the actual increase in length $\Delta L$ of the patch can be calculated using the following equation \cite{6,7,8}
\begin{align}
\frac{\Delta L}{h} = 0.412 \frac{(\epsilon_{\rm{reff}} + 0.3)(\frac{W}{h} + 0.264)}{(\epsilon_{\rm{reff}} - 0.258)(\frac{W}{h} + 0.8)},
\end{align}
The length ($L$) of the patch is given by the following equation \cite{6,7,8,21}
\begin{align}
L = \frac{c}{2f_0\sqrt{\epsilon_{\rm{reff}}}}-2\Delta L,
\end{align}
After solving these equations, we come up with the optimal parameters given by Table I.
\begin{table}[H]
\caption{Antenna Design Parameters}
\centering
\begin{tabular}{p{25pt}|p{75pt}}
\hline\hline\\
Symbol & \centerline{\textsc{Measure}}\\
\hline\\
$\bold{W}$& 17.30 mm \\
$\bold{L}$& 10.402 mm \\
$\bold{h}$& 1.6 mm \\ 
$\bold{\epsilon_r}$& 5 \\
$\bold{x_{feed}}$& 2.8 mm \\ 
\hline\hline
\end{tabular}
\end{table}
To realize a circular polarization, the feeding point of the microstrip should be located at the diagonal of the patch. The y-coordinate of the feeding point should be given by \cite{9}
\begin{align}
y_{\rm{feed}} = \frac{W}{L}x_{\rm{feed}},
\end{align}
A typical material with relative permittivity roughly equal to 5 can be FR-4 which is a material composed of a woven fiber glass cloth with an epoxy resin binder that is flame resistant. Other materials that can be used are the glass, Pyrex glass, mica, and porcelain.\\
Fig.~1 presents the patch antenna elements along with the materials required for the fabrication.
\begin{figure}[H]
\centering
\includegraphics[width=0.5\textwidth, bb =0 0 560 420]{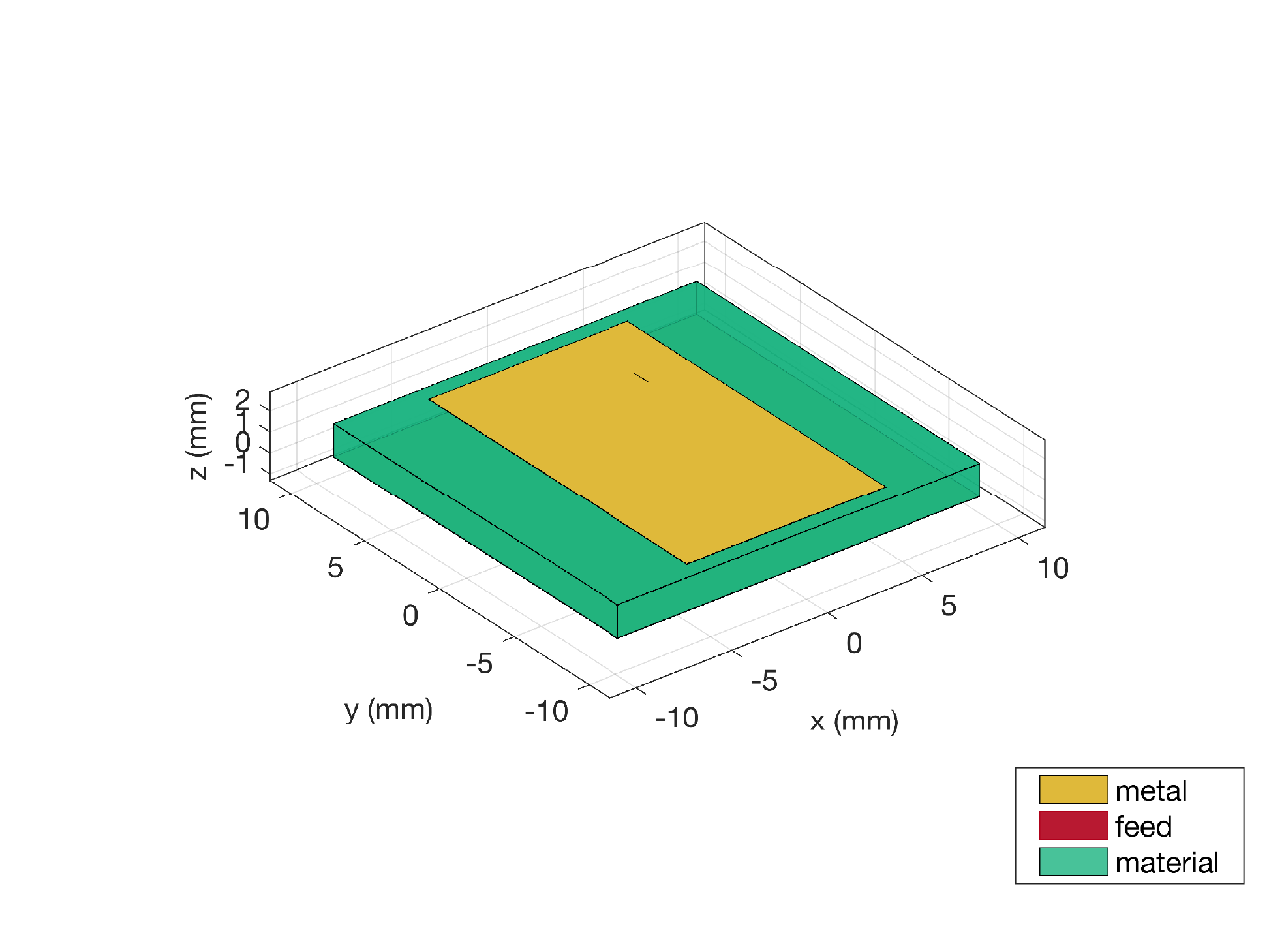}
\captionof{figure}{Patch Antenna Element.}
\label{fig1}
\end{figure}

\begin{figure}[H]
\centering
\includegraphics[width=0.5\textwidth,bb =0 0 560 420]{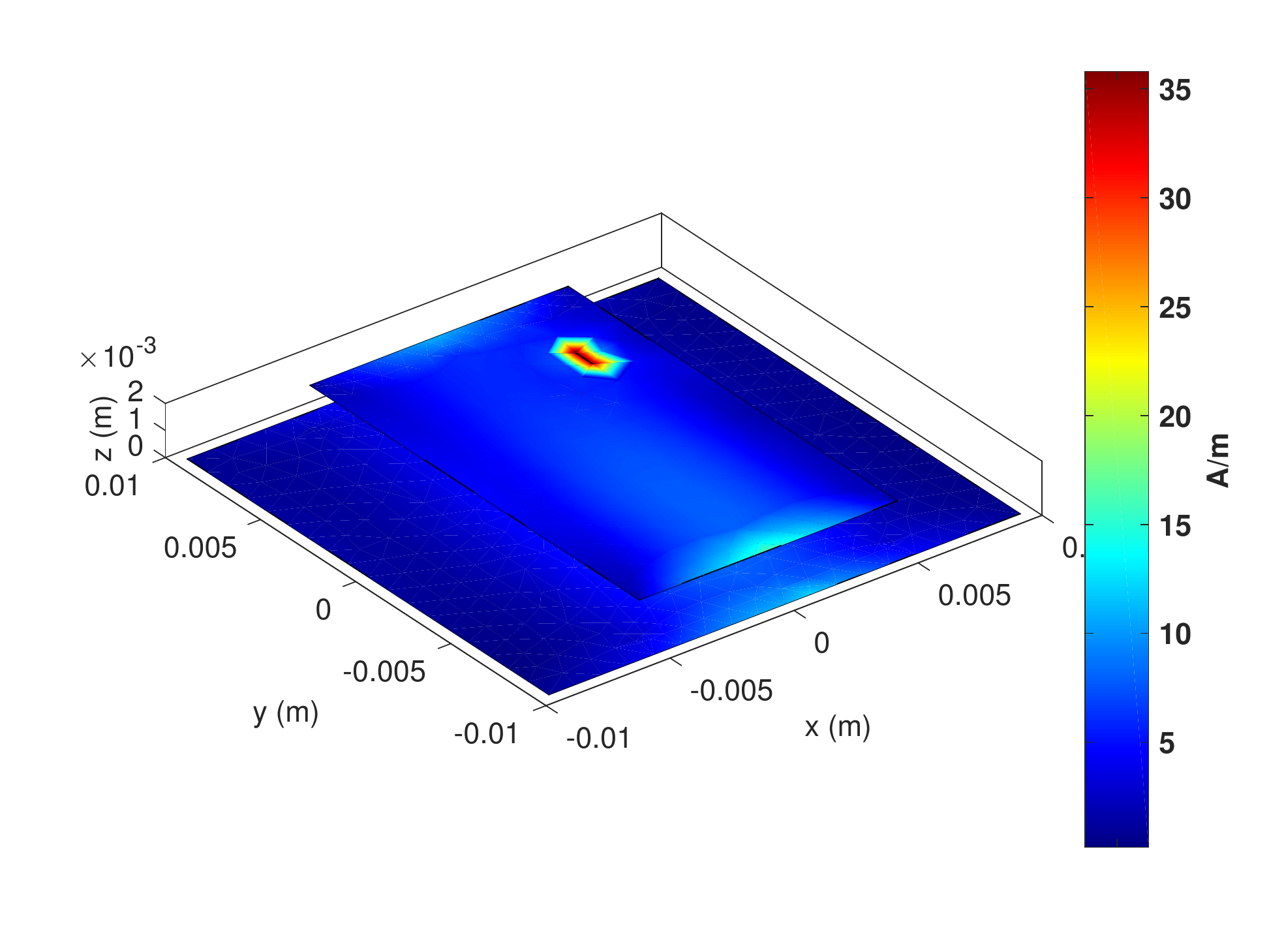}
\captionof{figure}{Current Distribution.}
\label{fig1}
\end{figure}
Fig.~2 shows that the current is maximum at the feed point and zero at the end of the patch. Given that the voltage is out of phase with the current, it is at a peak at the end of the patch, and a half-wavelength away at the start of the patch, it has equal magnitude but out of phase. It is this voltage, out of phase, which produces fringing fields that coherently add in phase and produce radiation.\\
Fig.~3 shows the electromagnetic fields radiated by the microstrip. The electric field is zero at the center of the patch, maximum (positive) on one side, and minimum (negative) on the opposite side. These minima and maxima continuously change side like the phase of the RF signal. The electric field does not stop abruptly near the patch's edges like it would in a cavity: the field extends beyond the outer periphery. These field extensions are the fringing fields and cause the patch to radiate. 
\begin{figure}[H]
\centering
\includegraphics[width=0.5\textwidth,bb =0 0 557 504]{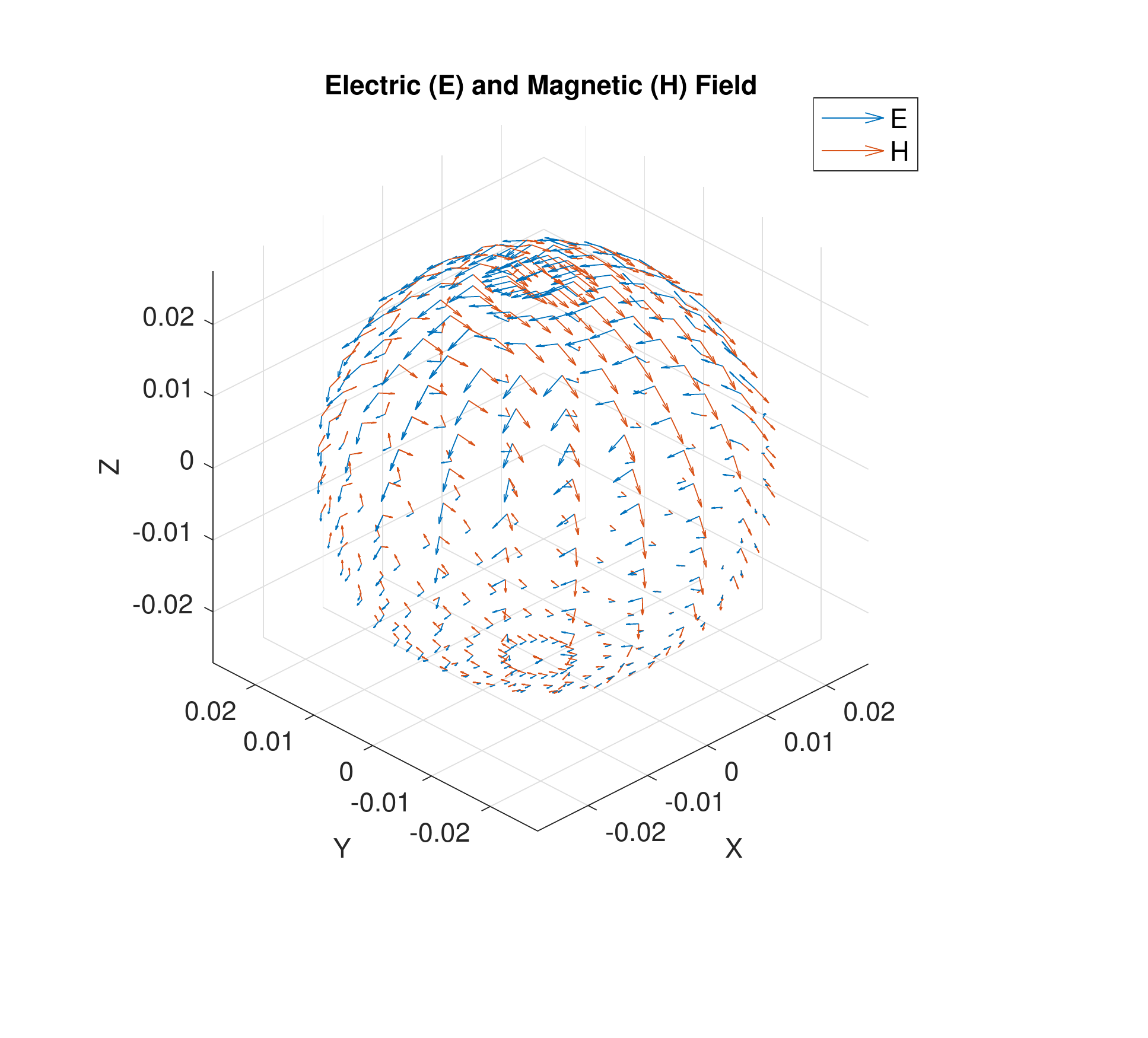}
\captionof{figure}{Electromagnetic Fields radiated by the microstrip.}
\label{fig1}
\end{figure}
Fig.~4 presents the three-dimentional radiation pattern of the patch. The radiated power is pronounced in particular directions and small in other ones. The pattern shows that the maximum radiated power is along with the direction perpendicular to the surface of the patch and in the upper half of the hemisphere. In the lower half of the hemisphere, the pattern also shows the existance of a back lobe with a low level compared to the main lobe level. 
\begin{figure}[H]
\centering
\includegraphics[width=0.5\textwidth,bb=0 0 560 420]{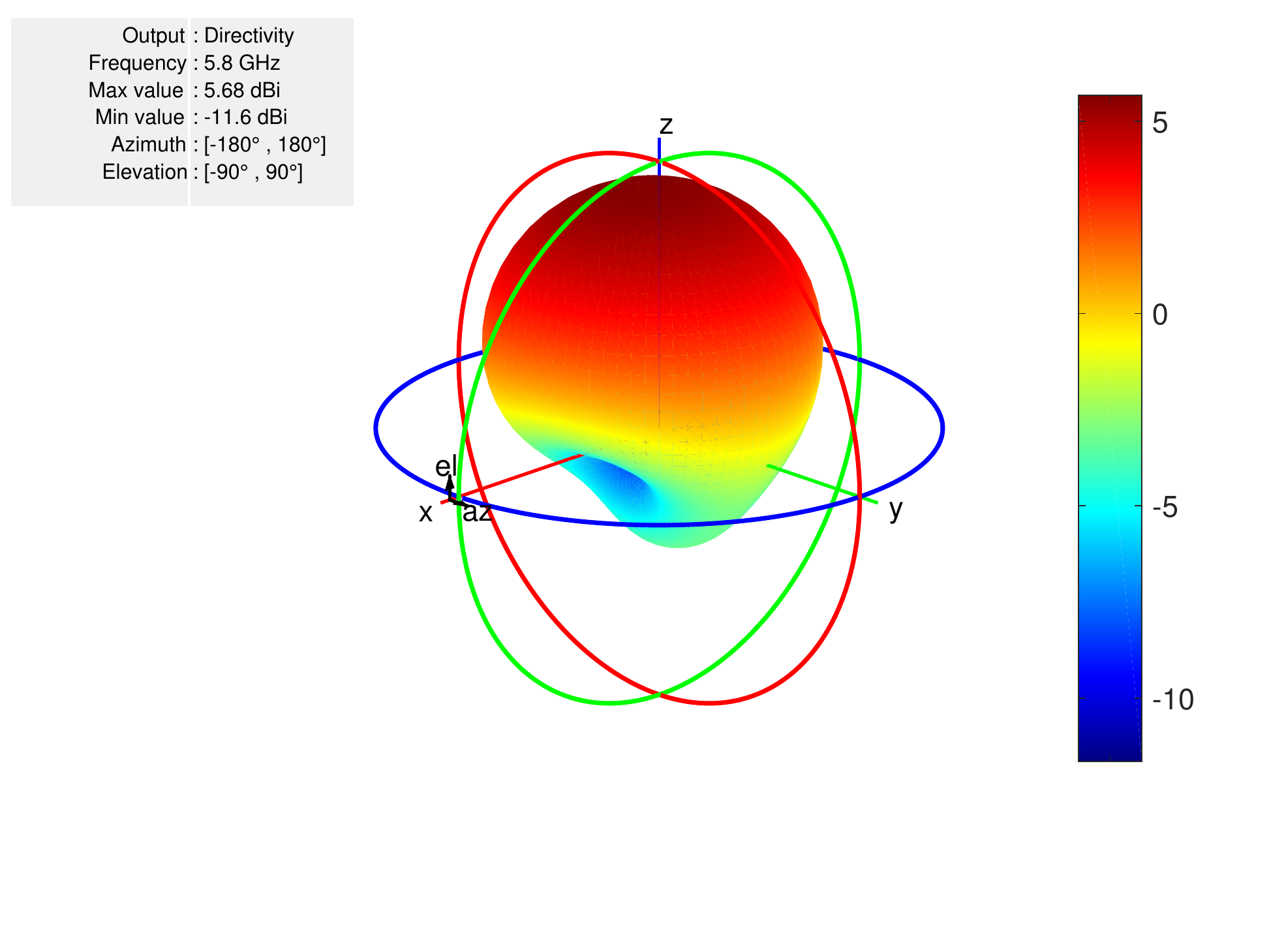}
\captionof{figure}{3D Radiation Pattern.}
\label{fig1}
\end{figure}
Fig.~5 illustrates the antenna directivity at the elevation plane. The pattern at this plane is obtained by slicing through the y-z plane. The figure shows that there is one main lobe radiated out from the front of the patch. The main lobe has a level of 5.62 dB located at 88$^\circ$, a half power beam width (HPBW) of 95.6$^\circ$, and a first null beam width (FNBW) of 236$^\circ$. The pattern also shows a back lobe with a level of -3.2 dB located at -92$^\circ$. The front to back ratio at this plane is about 8.83 dB.
\begin{figure}[H]
\centering
\includegraphics[width=0.5\textwidth,bb= 0 0 1120 840]{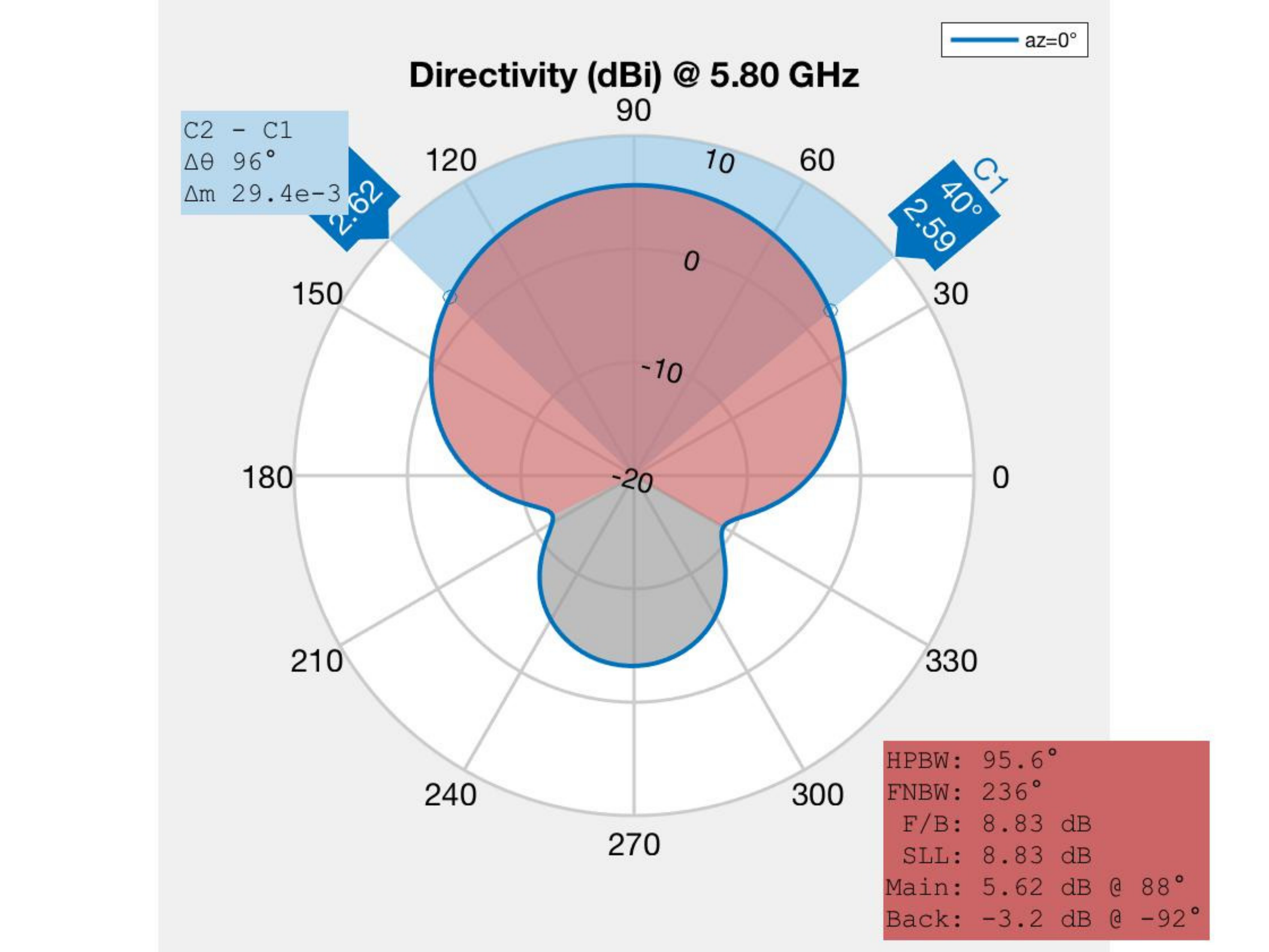}
\captionof{figure}{Directivity at the Elevation Plane.}
\label{fig1}
\end{figure}
\begin{figure}[H]
\centering
\includegraphics[width=0.5\textwidth,bb= 0 0 1120 840]{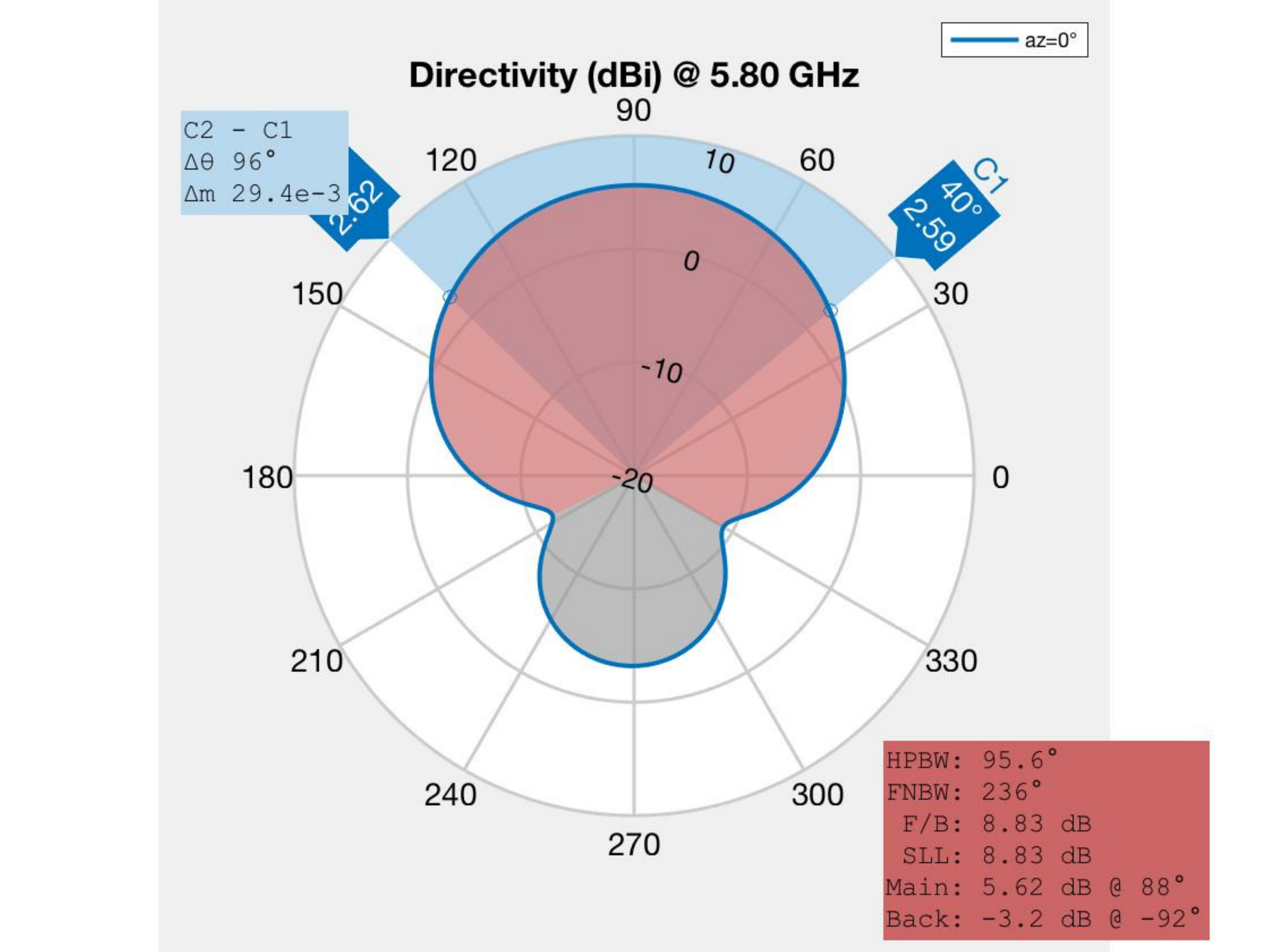}
\captionof{figure}{Directivity at the Azimuthal Plane.}
\label{fig1}
\end{figure}
Fig.~6 illustrates the antenna directivity at the azimuthal plane. The pattern at this plane is obtained by slicing through the x-y plane. The figure shows that there is one main lobe which has a level of -1.12 dB located at -84$^\circ$, a HPBW of 148$^\circ$, and a FNBW of 187$^\circ$. The pattern also shows a back lobe with a level of -1.47 dB located at -264$^\circ$. The front to back ratio at this plane is about 0.354 dB.
\begin{figure*}
\begin{center}
\begin{subfigure}[b]{.3\linewidth}
  \centering
  \setlength\fheight{0.6\textwidth}
  \setlength\fwidth{1.0\textwidth}
%
%
\begin{tikzpicture}

\begin{axis}[%
width=0.951\fwidth,
height=\fheight,
at={(0\fwidth,0\fheight)},
scale only axis,
xmin=5.3,
xmax=6.3,
xlabel style={font=\color{white!15!black}},
xlabel={Frequency (GHz)},
ymin=1,
ymax=7,
ylabel style={font=\color{white!15!black}},
ylabel={Magnitude (dB)},
axis background/.style={fill=white},
title style={font=\bfseries},
title={},
legend style={legend cell align=left, align=left, draw=white!15!black}
]
\addplot [color=blue, line width=.7pt]
  table[row sep=crcr]{%
5.3	6.57220310313625\\
5.32040816326531	6.16163539546666\\
5.34081632653061	5.77176663291775\\
5.36122448979592	5.40198503195082\\
5.38163265306123	5.05168128866625\\
5.40204081632653	4.72025110571731\\
5.42244897959184	4.40709744757029\\
5.44285714285714	4.11163265679069\\
5.46326530612245	3.83328059646465\\
5.48367346938776	3.57147902799435\\
5.50408163265306	3.32568250324298\\
5.52448979591837	3.09536615838414\\
5.54489795918367	2.88003096725819\\
5.56530612244898	2.67921127833048\\
5.58571428571429	2.49248589016052\\
5.60612244897959	2.31949461609044\\
5.6265306122449	2.15996341881044\\
5.6469387755102	2.01374299353199\\
5.66734693877551	1.8808683931876\\
5.68775510204082	1.76165074672981\\
5.70816326530612	1.65681413203193\\
5.72857142857143	1.56768195927974\\
5.74897959183674	1.49637014939818\\
5.76938775510204	1.44580883958914\\
5.78979591836735	1.41921383130512\\
5.81020408163265	1.41876516099977\\
5.83061224489796	1.44425215568578\\
5.85102040816327	1.49311452656093\\
5.87142857142857	1.56183494686525\\
5.89183673469388	1.64723548644337\\
5.91224489795918	1.74696828817341\\
5.93265306122449	1.85947025644957\\
5.9530612244898	1.98375899223557\\
5.9734693877551	2.11924156023546\\
5.99387755102041	2.26557485426929\\
6.01428571428571	2.42257151096673\\
6.03469387755102	2.59013763376864\\
6.05510204081633	2.76823097259417\\
6.07551020408163	2.95683178952497\\
6.09591836734694	3.15592138449915\\
6.11632653061224	3.36546505618378\\
6.13673469387755	3.58539739073265\\
6.15714285714286	3.81560845108421\\
6.17755102040816	4.05592985081415\\
6.19795918367347	4.30611995169959\\
6.21836734693878	4.56584758167212\\
6.23877551020408	4.83467377759684\\
6.25918367346939	5.11203114823473\\
6.27959183673469	5.39720053314997\\
6.3	5.68928474938275\\
};

\end{axis}
\end{tikzpicture}%
  \caption{VSWR.}
  \label{fig:mse_omegadot}
\end{subfigure}\hspace*{0.5cm}
\begin{subfigure}[b]{.3\linewidth}
  \centering
  \setlength\fheight{0.6\textwidth}
  \setlength\fwidth{1.0\textwidth}
%
%
\definecolor{mycolor1}{rgb}{0.00000,0.44700,0.74100}%
\begin{tikzpicture}

\begin{axis}[%
width=0.951\fwidth,
height=\fheight,
at={(0\fwidth,0\fheight)},
scale only axis,
xmin=5.300000000,
xmax=6.300000000,
xlabel style={font=\color{white!15!black}},
xlabel={Frequency (GHz)},
ymin=-16,
ymax=-2,
ylabel style={font=\color{white!15!black}},
ylabel={},
axis background/.style={fill=white},
legend style={legend cell align=left, align=left, draw=white!15!black}
]
\addplot [color=mycolor1, line width=.7pt]
  table[row sep=crcr]{%
5.300000000	-2.66390632784752\\
5.32040816326531	-2.84449766213329\\
5.34081632653061	-3.04045574280196\\
5.36122448979592	-3.25332185848684\\
5.38163265306122	-3.48481544183723\\
5.40204081632653	-3.73685678582952\\
5.42244897959184	-4.01159282836942\\
5.44285714285714	-4.31142630542978\\
5.46326530612245	-4.63904842956903\\
5.48367346938776	-4.99747490037805\\
5.50408163265306	-5.3900842865377\\
5.52448979591837	-5.82065623493743\\
5.54489795918367	-6.2934037782835\\
5.56530612244898	-6.81298770728218\\
5.58571428571429	-7.3844885235861\\
5.60612244897959	-8.01328692165192\\
5.6265306122449	-8.70475523482266\\
5.6469387755102	-9.463566962228\\
5.66734693877551	-10.292247833643\\
5.68775510204082	-11.1882576456367\\
5.70816326530612	-12.1383736835151\\
5.72857142857143	-13.1087227398783\\
5.74897959183673	-14.0300664943151\\
5.76938775510204	-14.7854766954399\\
5.78979591836735	-15.2247730745702\\
5.81020408163265	-15.2324632201561\\
5.83061224489796	-14.8103292605872\\
5.85102040816327	-14.0758886539483\\
5.87142857142857	-13.1788479265435\\
5.89183673469388	-12.2346051063633\\
5.91224489795918	-11.3110296235015\\
5.93265306122449	-10.4410946350403\\
5.9530612244898	-9.63750051451115\\
5.9734693877551	-8.90250360921502\\
5.99387755102041	-8.23343611952677\\
6.01428571428571	-7.62556847878923\\
6.03469387755102	-7.07352764278513\\
6.05510204081633	-6.57197042842404\\
6.07551020408163	-6.11588186301942\\
6.09591836734694	-5.70068806679854\\
6.11632653061224	-5.3222795879007\\
6.13673469387755	-4.9769934668297\\
6.15714285714286	-4.66157815735187\\
6.17755102040816	-4.37315317790005\\
6.19795918367347	-4.10916914151528\\
6.21836734693878	-3.86737066819692\\
6.23877551020408	-3.64576311118083\\
6.25918367346939	-3.44258326188052\\
6.27959183673469	-3.25627385625153\\
6.300000000	-3.08546156008456\\
};

\end{axis}
\end{tikzpicture}%
  \caption{Parameter $|S_{11}|$.}
  \label{fig:mse_p}
\end{subfigure}\hspace*{0.5cm}
\begin{subfigure}[b]{.3\linewidth}
  \centering
  \setlength\fheight{0.6\textwidth}
  \setlength\fwidth{1.0\textwidth}
%
%
\begin{tikzpicture}

\begin{axis}[%
width=0.951\fwidth,
height=\fheight,
at={(0\fwidth,0\fheight)},
scale only axis,
xmin=5.3,
xmax=6.3,
xlabel style={font=\color{white!15!black}},
xlabel={Frequency (GHz)},
ymin=2,
ymax=16,
ylabel style={font=\color{white!15!black}},
ylabel={},
axis background/.style={fill=white},
title style={font=\bfseries},
title={},
legend style={legend cell align=left, align=left, draw=white!15!black}
]
\addplot [color=blue, line width=.7pt]
  table[row sep=crcr]{%
5.3	2.66390632784752\\
5.32040816326531	2.84449766213329\\
5.34081632653061	3.04045574280196\\
5.36122448979592	3.25332185848684\\
5.38163265306123	3.48481544183723\\
5.40204081632653	3.73685678582952\\
5.42244897959184	4.01159282836942\\
5.44285714285714	4.31142630542978\\
5.46326530612245	4.63904842956903\\
5.48367346938776	4.99747490037805\\
5.50408163265306	5.3900842865377\\
5.52448979591837	5.82065623493743\\
5.54489795918367	6.2934037782835\\
5.56530612244898	6.81298770728218\\
5.58571428571429	7.3844885235861\\
5.60612244897959	8.01328692165192\\
5.6265306122449	8.70475523482266\\
5.6469387755102	9.463566962228\\
5.66734693877551	10.292247833643\\
5.68775510204082	11.1882576456367\\
5.70816326530612	12.1383736835151\\
5.72857142857143	13.1087227398783\\
5.74897959183674	14.0300664943151\\
5.76938775510204	14.7854766954399\\
5.78979591836735	15.2247730745702\\
5.81020408163265	15.2324632201561\\
5.83061224489796	14.8103292605872\\
5.85102040816327	14.0758886539483\\
5.87142857142857	13.1788479265435\\
5.89183673469388	12.2346051063633\\
5.91224489795918	11.3110296235015\\
5.93265306122449	10.4410946350403\\
5.9530612244898	9.63750051451115\\
5.9734693877551	8.90250360921502\\
5.99387755102041	8.23343611952677\\
6.01428571428571	7.62556847878923\\
6.03469387755102	7.07352764278513\\
6.05510204081633	6.57197042842404\\
6.07551020408163	6.11588186301942\\
6.09591836734694	5.70068806679854\\
6.11632653061224	5.3222795879007\\
6.13673469387755	4.9769934668297\\
6.15714285714286	4.66157815735187\\
6.17755102040816	4.37315317790005\\
6.19795918367347	4.10916914151528\\
6.21836734693878	3.86737066819692\\
6.23877551020408	3.64576311118083\\
6.25918367346939	3.44258326188052\\
6.27959183673469	3.25627385625153\\
6.3	3.08546156008456\\
};

\end{axis}
\end{tikzpicture}%
  \caption{Return Loss.}
  \label{fig:mse_p}
\end{subfigure}\hfill
\end{center}
\caption{Magnitudes variations of VSWR, $|S_{11}|$, and return loss with respect to the frequency.}
\end{figure*}
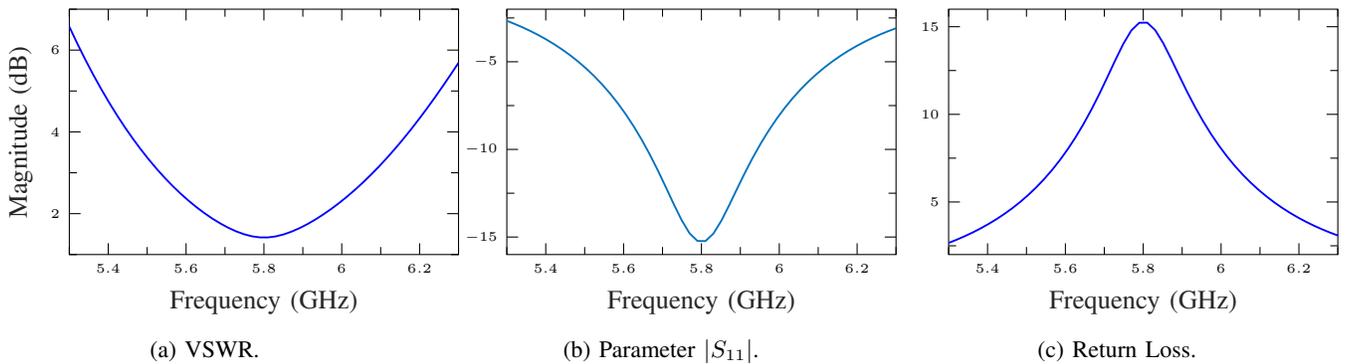\\
Fig.~7.a shows the variations of the voltage standing wave ratio (VSWR) with respect to the frequency. VSWR is a measure of how much power is delivered to a device as opposed to the amount of power that is reflected from the load. According to Fig.~7.a, the VSWR reaches a minimum of 1.3 dB at the resonant frequency 5.8 GHz which means that the source and the load impedance are matched at this particular frequency.\\
Fig.~7.b illustrates the dependence of $|S_{11}|$ with respect to the frequency. The minimum value of $|S_{11}|$ is -15.5 dB and  occurs at 5.8 GHz. The bandwidth is given by the frequency range where the magnitude of $|S_{11}|$ is below -6 dB. We observe that the bandwidth is roughly $BW$ $\cong$ 540 MHz. Note that the parameter $|S_{11}|$ is the opposite of the return loss given by Fig.~7.c. The maximum value of the return loss is 15.5 dB and achieved at 5.8 GHz, which means that the reflected power is minimum compared to the delivered power.  
\begin{figure}[H]
\centering
\setlength\fheight{5cm}
\setlength\fwidth{8cm}
%
%
\begin{tikzpicture}

\begin{axis}[%
width=0.951\fwidth,
height=\fheight,
at={(0\fwidth,0\fheight)},
scale only axis,
xmin=5.3,
xmax=6.3,
xlabel style={font=\color{white!15!black}},
xlabel={Frequency (GHz)},
ymin=5,
ymax=50,
ylabel style={font=\color{white!15!black}},
ylabel={Impedance (ohms)},
axis background/.style={fill=white},
title style={font=\bfseries},
title={},
legend style={legend cell align=left, align=left, draw=white!15!black}
]
\addplot [color=blue, line width=.7pt]
  table[row sep=crcr]{%
5.3	13.547778438417\\
5.32040816326531	14.7605631091141\\
5.34081632653061	16.0968425108821\\
5.36122448979592	17.566982745223\\
5.38163265306123	19.1806227514024\\
5.40204081632653	20.945792237085\\
5.42244897959184	22.8676846158433\\
5.44285714285714	24.9470223126772\\
5.46326530612245	27.177983339975\\
5.48367346938776	29.5457280806816\\
5.50408163265306	32.0236891186215\\
5.52448979591837	34.5709728907943\\
5.54489795918367	37.1304555909903\\
5.56530612244898	39.6283804495957\\
5.58571428571429	41.9763660552637\\
5.60612244897959	44.0765602465397\\
5.6265306122449	45.8300921870509\\
5.6469387755102	47.148004637605\\
5.66734693877551	47.9627620082589\\
5.68775510204082	48.2377257824838\\
5.70816326530612	47.9721476430354\\
5.72857142857143	47.2003650982664\\
5.74897959183674	45.985569088149\\
5.76938775510204	44.4099963501567\\
5.78979591836735	42.5640697493333\\
5.81020408163265	40.5367407473246\\
5.83061224489796	38.4084122917765\\
5.85102040816327	36.2468318343823\\
5.87142857142857	34.1056061506432\\
5.89183673469388	32.024625391312\\
5.91224489795918	30.0316332002897\\
5.93265306122449	28.1443109778678\\
5.9530612244898	26.3724380318379\\
5.9734693877551	24.7198706993665\\
5.99387755102041	23.1862210893003\\
6.01428571428571	21.7682056880776\\
6.03469387755102	20.460684462362\\
6.05510204081633	19.2574342773961\\
6.07551020408163	18.1517069002154\\
6.09591836734694	17.1366193681946\\
6.11632653061224	16.2054179697429\\
6.13673469387755	15.3516494377273\\
6.15714285714286	14.5692656732896\\
6.17755102040816	13.8526820380668\\
6.19795918367347	13.1968041490249\\
6.21836734693878	12.5970341376732\\
6.23877551020408	12.0492643219802\\
6.25918367346939	11.5498640051687\\
6.27959183673469	11.0956635212413\\
6.3	10.6839385082475\\
};
\addlegendentry{Resistance}

\addplot [color=red, line width=.7pt]
  table[row sep=crcr]{%
5.3	43.2604212752744\\
5.32040816326531	44.1516041418484\\
5.34081632653061	45.0077466856926\\
5.36122448979592	45.8129619932114\\
5.38163265306123	46.5474737798075\\
5.40204081632653	47.1870249604159\\
5.42244897959184	47.7023718912128\\
5.44285714285714	48.058993993445\\
5.46326530612245	48.2172100278287\\
5.48367346938776	48.1329586863474\\
5.50408163265306	47.7595526895048\\
5.52448979591837	47.0507160797198\\
5.54489795918367	45.9651106598312\\
5.56530612244898	44.4722920598169\\
5.58571428571429	42.5595811400379\\
5.60612244897959	40.2387485955113\\
5.6265306122449	37.5508789653481\\
5.6469387755102	34.5676122390203\\
5.66734693877551	31.3874504254053\\
5.68775510204082	28.1270139369915\\
5.70816326530612	24.9086798207207\\
5.72857142857143	21.8472562730543\\
5.74897959183674	19.0386378410225\\
5.76938775510204	16.5526096974781\\
5.78979591836735	14.4305530400272\\
5.81020408163265	12.6874253683205\\
5.83061224489796	11.316556680975\\
5.85102040816327	10.2956393707806\\
5.87142857142857	9.59260811306385\\
5.89183673469388	9.17061620955876\\
5.91224489795918	8.99178946825269\\
5.93265306122449	9.01977021341852\\
5.9530612244898	9.22124110086088\\
5.9734693877551	9.56667809197065\\
5.99387755102041	10.0305714856663\\
6.01428571428571	10.5913113891284\\
6.03469387755102	11.2308838816789\\
6.05510204081633	11.9344786203202\\
6.07551020408163	12.6900726243169\\
6.09591836734694	13.4880286933129\\
6.11632653061224	14.3207289761517\\
6.13673469387755	15.1822526825864\\
6.15714285714286	16.0680999352021\\
6.17755102040816	16.974959795682\\
6.19795918367347	17.9005185767096\\
6.21836734693878	18.8433037900797\\
6.23877551020408	19.8025590456483\\
6.25918367346939	20.778145590098\\
6.27959183673469	21.7704666621699\\
6.3	22.7804115080132\\
};
\addlegendentry{Reactance}

\end{axis}
\end{tikzpicture}%
    \caption{Input impedance.}
\end{figure}
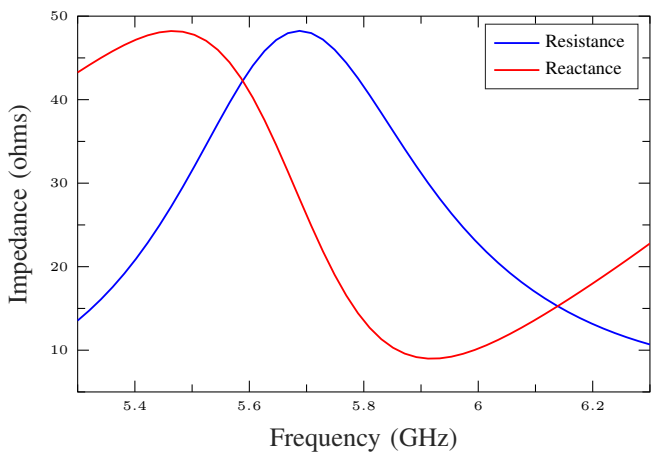
Fig.~8 presents the variations of real (resistance) and imaginary (reactance) parts of the input impedance with respect to the frequency. The resistance reaches a peak value of 47 $\Omega$ while the reactance gets a maximum of 47 $\Omega$ and a minimum of 9 $\Omega$.

\section{One-Dimensional Simulation}
\subsection{Yee Algorithm}
The FDTD algorithm as first proposed by Kane Yee in 1966 employs second-order central differences. The algorithm can be summarized as follows
\begin{algorithm}
\caption{Yee Algorithm}\label{euclid}
\begin{algorithmic}[1]
\State Replace all the derivatives in Ampere’s and Faraday’s laws with finite differences. Discretize space and time so that the electric and magnetic fields are staggered in both space and time.
\State Solve the resulting difference equations to obtain 'update equations' that express the (unknown) future fields in terms of (known) past fields.
\State Evaluate the magnetic fields one time-step into the future so they are now known (effectively they become past fields).
\State Evaluate the electric fields one time-step into the future so they are now known (effectively they become past fields).
\State Repeat the previous two steps until the fields have been obtained over the desired duration.
\end{algorithmic}
\end{algorithm}\\
Table V summarizes the main electromagnetic parameters used in the FDTD analysis.
\begin{table}[H]
\caption{Electromagnetic Parameters}
\centering
\begin{tabular}{p{25pt}|p{75pt}}
\hline\hline\\
Symbol & \centerline{\textsc{Quantity}}\\
\hline\\
$\bold{B}$& magnetic flux density \\
$\bold{H}$& magnetic field strength \\
$\bold{D}$& electric field density \\ 
$\bold{E}$& electric field strength \\
$\bold{\epsilon_0}$& permittivity of free space \\
$\bold{\mu_0}$& permeability of free space \\
$\bold{\epsilon_r}$& relative permittivity \\
$\bold{\mu_r}$& relative permeability \\
$\bold{\sigma}$& conductivity \\
$\bold{\eta}$& intrinsic impedance \\
$\bold{c}$& speed of light \\
\hline\hline
\end{tabular}
\end{table}
\subsection{FDTD Formulation}
Considering Maxwell's curl equations in free space as follows
\begin{align}\label{eq:1}
\frac{\partial \textbf{\textit{E}}}{\partial t} = \frac{1}{\epsilon_0} \nabla \times \textbf{\textit{H}},
\end{align}
\begin{align}\label{eq:2}
\frac{\partial \textbf{\textit{H}}}{\partial t} = -\frac{1}{\mu_0} \nabla \times \textbf{\textit{E}},
\end{align}
Given that one dimension is assumed, the components $E_x$ and $H_y$ are only considered to facilitate the calculus and the other components are set to zero. Given that the wave propagates in $z$ direction, the curl equations are reduced as follows
\begin{align}\label{eq:3}
\frac{\partial E_x}{\partial t} = -\frac{1}{\epsilon_0} \frac{\partial H_y}{\partial z},
\end{align}
\begin{align}\label{eq:4}
\frac{\partial H_y}{\partial t} = -\frac{1}{\mu_0} \frac{\partial E_x}{\partial z},
\end{align}
To discretize these equations in time and space, we should apply the central difference method for temporal and spatial derivatives. The above aligns can be written as follows
\begin{align}\label{eq:5}
\frac{E_x^{n+\frac{1}{2}}(k) - E_x^{n-\frac{1}{2}}(k)}{\Delta t} = -\frac{1}{\epsilon_0} \frac{H_y^{n}\left(k+\frac{1}{2}\right) - H_y^{n}\left(k-\frac{1}{2}\right)}{\Delta x},
\end{align}

\begin{align}\label{eq:6}
\frac{H_y^{n+1}\left(k+\frac{1}{2}\right) - H_y^{n}\left(k+\frac{1}{2}\right)}{\Delta t} = -\frac{E_x^{n+\frac{1}{2}}(k+1) - E_x^{n+\frac{1}{2}}(k)}{\mu_0\Delta x},
\end{align}
where $\Delta t$ and $\Delta x$ are the time and the space step, respectively. $n$ and $k$ are the number of the time and space steps, respectively. An instant $t$ is given by $t = n \Delta t$, while a distance $z$ is given by $z = k \Delta x$. The discretized equations can be formulated into an iterative algorithm as follows
\begin{align}\label{eq:7}
E_x^{n+\frac{1}{2}}(k)  =& E_x^{n-\frac{1}{2}}(k) -\frac{\Delta t}{\epsilon_0\Delta x}\\& \times~ \left[ H_y^{n}\left(k+\frac{1}{2}\right) - H_y^{n}\left(k-\frac{1}{2}\right)\right],
\end{align}
\begin{align}\label{eq:8}
\begin{split}
H_y^{n+1}\left(k+\frac{1}{2}\right) =& H_y^{n}\left(k+\frac{1}{2}\right) -\frac{\Delta t}{\mu_0\Delta x} \\& \times~ \left[E_x^{n+\frac{1}{2}}(k+1) - E_x^{n+\frac{1}{2}}(k)\right],
\end{split}
\end{align}
Note that $E_x$ and $H_y$ differ by a large order of magnitude because of $\epsilon_0$ and $\mu_0$. To fix this problem, we will roughly normalize the magnitude of the electric field so that it becomes comparable to the amplitude of the magnetic field. The new expression of the electric field can be obtained by \cite[Eq.~(1.5)]{sul}
\begin{align}\label{eq:9}
\widetilde{E} = \sqrt{\frac{\epsilon_0}{\mu_0}}E,
\end{align}
As we mentioned earlier, the space is equally discretized into cells of the same size. For a given cell of size $\Delta x$, the time step is given by
\begin{align}\label{eq:10}
\Delta t = \frac{\Delta x}{2 c},
\end{align}
where $c$ is the speed of light.\\ 
In this work, we create a Gaussian pulse in the middle and once the pulse is launched at $t_0$, the wave propagates in the two directions (positive and negative $z$ directions) until it hits the boundaries, then it reflects back. Given that the boundaries are absorbing, the waves will not be reflected back. The first boundary can be modelled as follows
\begin{align}\label{eq:11}
E_x^n(0) = E_x^{n-2}(1),
\end{align}
The electric field can be given at the second boundary as follows
\begin{align}\label{eq:12}
E_x^n(\text{KE}) = E_x^{n-2}(\text{KE}-1),
\end{align}
where KE is the length of the propagation medium.
\subsection{Analytic Solution}
Consider a Gaussian pulse as follows
\begin{equation}\label{eq:13}
f(t) = e^{-\sigma t^2},
\end{equation}
The electromagnetic field is assumed to propagate in both positive and negative $z$ directions. If the time pulse is Gaussian, the spectral density of this pulse is also Gaussian. The temporal Fourier transform of the initial pulse is given by
\begin{equation}\label{eq:14}
F(\omega) = \sqrt{\frac{\pi}{\sigma}}e^{-\frac{\omega^2}{4\sigma}},
\end{equation}
Note that the pulse in frequency domain should be multiplied by a carrier or a propagator factor $e^{-ikz}$ to move in the positive $z$ direction and a carrier $e^{+ikz}$ to move in the negative $z$ direction. To get the expression of the electric field in time and space (positive $z$ direction), the next step is to take the inverse of the Fourier transform of the spectral pulse multiplied by the carrier as follows
\begin{equation}\label{eq:15}
E^{+}(z,t) = \frac{1}{2\pi}\int\limits_0^\infty F(\omega)~e^{-ikz}e^{+i\omega t}d\omega,
\end{equation}
The electric field $E^{-}(z,t)$ propagating in the negative $z$ direction can be derived by replacing the propagating factor in Eq.~(21) by $e^{+ikz}$. Finally, the total electric field in time and space is given by
\begin{equation}\label{eq:16}
 E(z,t) = E^{+}(z,t) + E^{-}(z,t),   
\end{equation}
A Gaussian pulse is created at the instant $t_0$ in the middle of the space problem taking the following form:
\begin{equation}\label{eq:17}
f(t) = e^{\frac{(t_0 - t)^2}{2\rho^2}},
\end{equation}
The Fourrier transform of the above pulse is given by
\begin{align}
F(\omega) = \rho~e^{-\frac{(\omega \rho)^2}{2}}e^{i\omega t_0},
\end{align}
The electric field in the positive z direction can be given by
\begin{align}
E^{+}(z,t) = \rho^2~e^{\frac{(t_0 - t)^2}{2\rho^2}} e^{-ikz},
\end{align}
The electric field in the negative z direction can be given by
\begin{align}
E^{-}(z,t) = \rho^2~e^{\frac{(t_0 - t)^2}{2\rho^2}} e^{+ikz},
\end{align}
\subsection{Numerical Simulations}
\begin{figure}[H]
\centering
\setlength\fheight{5cm}
\setlength\fwidth{8cm}
%
%
\begin{tikzpicture}

\begin{axis}[%
width=0.951\fwidth,
height=\fheight,
at={(0\fwidth,0\fheight)},
scale only axis,
xmin=0,
xmax=200,
xlabel style={font=\color{white!15!black}},
xlabel={FDTD cells},
ymin=-0.5,
ymax=1.5,
ylabel style={font=\color{white!15!black}},
ylabel={$\text{E}_\text{x}\text{ (V/m)}$},
axis background/.style={fill=white},
legend style={at={(0.5,0.97)}, anchor=north, legend cell align=left, align=left, draw=white!15!black}
]
\addplot [color=black, line width=0.7pt]
  table[row sep=crcr]{%
1	1.91547895196715e-29\\
2	1.2842313672676e-28\\
3	8.37424045460638e-28\\
4	5.3110922496791e-27\\
5	3.27611039093474e-26\\
6	1.96548382416367e-25\\
7	1.1468765822256e-24\\
8	6.50878851546785e-24\\
9	3.59269128537111e-23\\
10	1.92874984796392e-22\\
11	1.00708974473355e-21\\
12	5.11442373180375e-21\\
13	2.52616378092569e-20\\
14	1.21356366812579e-19\\
15	5.67021977492392e-19\\
16	2.57675710915498e-18\\
17	1.13889408651823e-17\\
18	4.89586526458611e-17\\
19	2.0469717131642e-16\\
20	8.32396967698111e-16\\
21	3.29219387117252e-15\\
22	1.26641655490942e-14\\
23	4.73809779628813e-14\\
24	1.72412093903551e-13\\
25	6.10193667760532e-13\\
26	2.10040929273226e-12\\
27	7.03196062084075e-12\\
28	2.28973484564555e-11\\
29	7.25153956932532e-11\\
30	2.23363143620316e-10\\
31	6.69158609129278e-10\\
32	1.94976778601724e-09\\
33	5.52551772036671e-09\\
34	1.52299797447126e-08\\
35	4.08283604114252e-08\\
36	1.0645371411076e-07\\
37	2.69957850336301e-07\\
38	6.65836146985732e-07\\
39	1.59725788283431e-06\\
40	3.72665317207867e-06\\
41	8.45666596870112e-06\\
42	1.86644691135205e-05\\
43	4.00652973929511e-05\\
44	8.36483472297277e-05\\
45	0.00016985667656141\\
46	0.000335462627902512\\
47	0.000644379820568608\\
48	0.0012038599948282\\
49	0.00218749111818289\\
50	0.0038659201394728\\
51	0.0066450112827414\\
52	0.0111089965382423\\
53	0.0180630134197813\\
54	0.0285655007845504\\
55	0.0439369336234074\\
56	0.0657285286165304\\
57	0.0956344448325386\\
58	0.135335283236613\\
59	0.186270463697701\\
60	0.249352208777296\\
61	0.32465246735835\\
62	0.411112290507187\\
63	0.506335616648101\\
64	0.606530659712633\\
65	0.706648277857716\\
66	0.800737402916808\\
67	0.882496902584595\\
68	0.945959468906765\\
69	0.986207116743916\\
70	1\\
71	0.986207116743916\\
72	0.945959468906765\\
73	0.882496902584595\\
74	0.800737402916808\\
75	0.706648277857716\\
76	0.606530659712633\\
77	0.506335616648101\\
78	0.411112290507188\\
79	0.32465246735835\\
80	0.249352208777297\\
81	0.186270463697704\\
82	0.135335283236625\\
83	0.095634444832586\\
84	0.0657285286167029\\
85	0.0439369336240176\\
86	0.0285655007866508\\
87	0.0180630134268132\\
88	0.0111089965611397\\
89	0.0066450113552568\\
90	0.00386592036283595\\
91	0.00218749178734149\\
92	0.00120386194459599\\
93	0.000644385346086328\\
94	0.000335477857882257\\
95	0.000169897504921822\\
96	8.37548009438384e-05\\
97	4.03352552432874e-05\\
98	1.93303052605063e-05\\
99	1.00539238515354e-05\\
100	7.45330634415734e-06\\
101	1.00539238515354e-05\\
102	1.93303052605063e-05\\
103	4.03352552432874e-05\\
104	8.37548009438384e-05\\
105	0.000169897504921822\\
106	0.000335477857882257\\
107	0.000644385346086328\\
108	0.00120386194459599\\
109	0.00218749178734149\\
110	0.00386592036283595\\
111	0.0066450113552568\\
112	0.0111089965611397\\
113	0.0180630134268132\\
114	0.0285655007866508\\
115	0.0439369336240176\\
116	0.0657285286167029\\
117	0.095634444832586\\
118	0.135335283236625\\
119	0.186270463697704\\
120	0.249352208777297\\
121	0.32465246735835\\
122	0.411112290507188\\
123	0.506335616648101\\
124	0.606530659712633\\
125	0.706648277857716\\
126	0.800737402916808\\
127	0.882496902584595\\
128	0.945959468906765\\
129	0.986207116743916\\
130	1\\
131	0.986207116743916\\
132	0.945959468906765\\
133	0.882496902584595\\
134	0.800737402916808\\
135	0.706648277857716\\
136	0.606530659712633\\
137	0.506335616648101\\
138	0.411112290507187\\
139	0.32465246735835\\
140	0.249352208777296\\
141	0.186270463697701\\
142	0.135335283236613\\
143	0.0956344448325386\\
144	0.0657285286165304\\
145	0.0439369336234074\\
146	0.0285655007845504\\
147	0.0180630134197813\\
148	0.0111089965382423\\
149	0.0066450112827414\\
150	0.0038659201394728\\
151	0.00218749111818289\\
152	0.0012038599948282\\
153	0.000644379820568608\\
154	0.000335462627902512\\
155	0.00016985667656141\\
156	8.36483472297277e-05\\
157	4.00652973929511e-05\\
158	1.86644691135205e-05\\
159	8.45666596870112e-06\\
160	3.72665317207867e-06\\
161	1.59725788283431e-06\\
162	6.65836146985732e-07\\
163	2.69957850336301e-07\\
164	1.0645371411076e-07\\
165	4.08283604114252e-08\\
166	1.52299797447126e-08\\
167	5.52551772036671e-09\\
168	1.94976778601724e-09\\
169	6.69158609129278e-10\\
170	2.23363143620316e-10\\
171	7.25153956932532e-11\\
172	2.28973484564555e-11\\
173	7.03196062084075e-12\\
174	2.10040929273226e-12\\
175	6.10193667760532e-13\\
176	1.72412093903551e-13\\
177	4.73809779628813e-14\\
178	1.26641655490942e-14\\
179	3.29219387117252e-15\\
180	8.32396967698111e-16\\
181	2.0469717131642e-16\\
182	4.89586526458611e-17\\
183	1.13889408651823e-17\\
184	2.57675710915498e-18\\
185	5.67021977492392e-19\\
186	1.21356366812579e-19\\
187	2.52616378092569e-20\\
188	5.11442373180375e-21\\
189	1.00708974473355e-21\\
190	1.92874984796392e-22\\
191	3.59269128537111e-23\\
192	6.50878851546785e-24\\
193	1.1468765822256e-24\\
194	1.96548382416367e-25\\
195	3.27611039093474e-26\\
196	5.3110922496791e-27\\
197	8.37424045460638e-28\\
198	1.2842313672676e-28\\
199	1.91547895196715e-29\\
200	2.77873901831711e-30\\
};
\addlegendentry{Analytic Solution}

\addplot [color=blue, dashed, line width=0.7pt]
  table[row sep=crcr]{%
1	1.91547895196715e-29\\
5	3.27611039093474e-26\\
9	3.59269128537111e-23\\
13	2.52616378092569e-20\\
17	1.13889408651823e-17\\
21	3.29219387117252e-15\\
25	6.10193667760532e-13\\
29	7.25153956932532e-11\\
33	5.52551772036671e-09\\
37	2.69957850336301e-07\\
41	8.45666596870112e-06\\
45	0.00016985667656141\\
49	0.00218749111818289\\
53	0.0180630134197813\\
57	0.0956344448325386\\
61	0.32465246735835\\
65	0.706648277857716\\
69	0.986207116743916\\
73	0.882496902584595\\
77	0.506335616648101\\
81	0.186270463697704\\
85	0.0439369336240176\\
89	0.0066450113552568\\
93	0.000644385346086328\\
97	4.03352552432874e-05\\
101	1.00539238515354e-05\\
105	0.000169897504921822\\
109	0.00218749178734149\\
113	0.0180630134268132\\
117	0.095634444832586\\
121	0.32465246735835\\
125	0.706648277857716\\
129	0.986207116743916\\
133	0.882496902584595\\
137	0.506335616648101\\
141	0.186270463697701\\
145	0.0439369336234074\\
149	0.0066450112827414\\
153	0.000644379820568608\\
157	4.00652973929511e-05\\
161	1.59725788283431e-06\\
165	4.08283604114252e-08\\
169	6.69158609129278e-10\\
173	7.03196062084075e-12\\
177	4.73809779628813e-14\\
181	2.0469717131642e-16\\
185	5.67021977492392e-19\\
189	1.00708974473355e-21\\
193	1.1468765822256e-24\\
197	8.37424045460638e-28\\
};
\addlegendentry{FDTD Method}

\end{axis}
\end{tikzpicture}%
    \caption{Electric field.}
\end{figure}
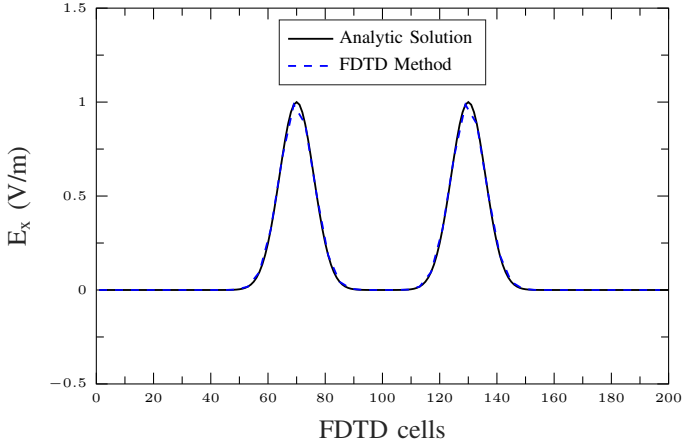
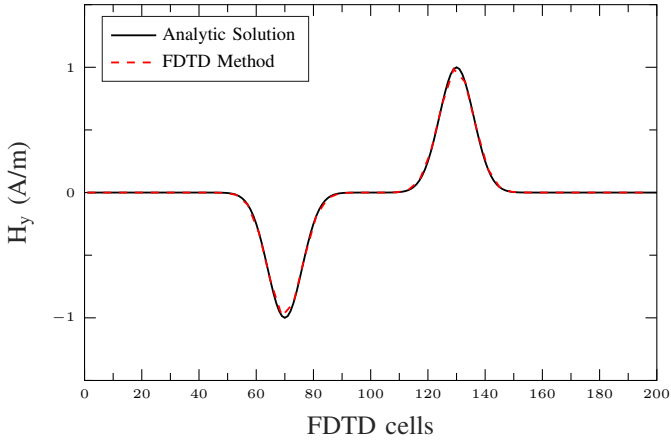
\begin{figure}[H]
\centering
\setlength\fheight{5cm}
\setlength\fwidth{8cm}
%
%
\begin{tikzpicture}

\begin{axis}[%
width=0.951\fwidth,
height=\fheight,
at={(0\fwidth,0\fheight)},
scale only axis,
xmin=0,
xmax=200,
xlabel style={font=\color{white!15!black}},
xlabel={FDTD cells},
ymin=-1.5,
ymax=1.5,
ylabel style={font=\color{white!15!black}},
ylabel={$\text{H}_\text{y}\text{ (A/m)}$},
axis background/.style={fill=white},
legend style={at={(0.03,0.97)}, anchor=north west, legend cell align=left, align=left, draw=white!15!black}
]
\addplot [color=black, line width=0.7pt]
  table[row sep=crcr]{%
1	-1.91547895196715e-29\\
2	-1.2842313672676e-28\\
3	-8.37424045460638e-28\\
4	-5.3110922496791e-27\\
5	-3.27611039093474e-26\\
6	-1.96548382416367e-25\\
7	-1.1468765822256e-24\\
8	-6.50878851546785e-24\\
9	-3.59269128537111e-23\\
10	-1.92874984796392e-22\\
11	-1.00708974473355e-21\\
12	-5.11442373180375e-21\\
13	-2.52616378092569e-20\\
14	-1.21356366812579e-19\\
15	-5.67021977492392e-19\\
16	-2.57675710915498e-18\\
17	-1.13889408651823e-17\\
18	-4.89586526458611e-17\\
19	-2.0469717131642e-16\\
20	-8.32396967698111e-16\\
21	-3.29219387117252e-15\\
22	-1.26641655490942e-14\\
23	-4.73809779628813e-14\\
24	-1.72412093903551e-13\\
25	-6.10193667760532e-13\\
26	-2.10040929273226e-12\\
27	-7.03196062084075e-12\\
28	-2.28973484564555e-11\\
29	-7.25153956932532e-11\\
30	-2.23363143620316e-10\\
31	-6.69158609129278e-10\\
32	-1.94976778601724e-09\\
33	-5.52551772036671e-09\\
34	-1.52299797447126e-08\\
35	-4.08283604114252e-08\\
36	-1.0645371411076e-07\\
37	-2.69957850336301e-07\\
38	-6.65836146985732e-07\\
39	-1.59725788283431e-06\\
40	-3.72665317207867e-06\\
41	-8.45666596870112e-06\\
42	-1.86644691135205e-05\\
43	-4.00652973929511e-05\\
44	-8.36483472297277e-05\\
45	-0.00016985667656141\\
46	-0.000335462627902512\\
47	-0.000644379820568608\\
48	-0.0012038599948282\\
49	-0.00218749111818289\\
50	-0.0038659201394728\\
51	-0.0066450112827414\\
52	-0.0111089965382423\\
53	-0.0180630134197813\\
54	-0.0285655007845504\\
55	-0.0439369336234074\\
56	-0.0657285286165304\\
57	-0.0956344448325386\\
58	-0.135335283236613\\
59	-0.186270463697701\\
60	-0.249352208777296\\
61	-0.32465246735835\\
62	-0.411112290507187\\
63	-0.506335616648101\\
64	-0.606530659712633\\
65	-0.706648277857716\\
66	-0.800737402916808\\
67	-0.882496902584595\\
68	-0.945959468906765\\
69	-0.986207116743916\\
70	-1\\
71	-0.986207116743916\\
72	-0.945959468906765\\
73	-0.882496902584595\\
74	-0.800737402916808\\
75	-0.706648277857716\\
76	-0.606530659712633\\
77	-0.506335616648101\\
78	-0.411112290507187\\
79	-0.32465246735835\\
80	-0.249352208777295\\
81	-0.186270463697698\\
82	-0.1353352832366\\
83	-0.0956344448324913\\
84	-0.065728528616358\\
85	-0.0439369336227972\\
86	-0.02856550078245\\
87	-0.0180630134127493\\
88	-0.011108996515345\\
89	-0.00664501121022601\\
90	-0.00386591991610966\\
91	-0.00218749044902428\\
92	-0.00120385804506042\\
93	-0.000644374295050887\\
94	-0.000335447397922767\\
95	-0.000169815848200999\\
96	-8.35418935156169e-05\\
97	-3.97953395426148e-05\\
98	-1.79986329665348e-05\\
99	-6.85940808586681e-06\\
100	0\\
101	6.85940808586681e-06\\
102	1.79986329665348e-05\\
103	3.97953395426148e-05\\
104	8.35418935156169e-05\\
105	0.000169815848200999\\
106	0.000335447397922767\\
107	0.000644374295050887\\
108	0.00120385804506042\\
109	0.00218749044902428\\
110	0.00386591991610966\\
111	0.00664501121022601\\
112	0.011108996515345\\
113	0.0180630134127493\\
114	0.02856550078245\\
115	0.0439369336227972\\
116	0.065728528616358\\
117	0.0956344448324913\\
118	0.1353352832366\\
119	0.186270463697698\\
120	0.249352208777295\\
121	0.32465246735835\\
122	0.411112290507187\\
123	0.506335616648101\\
124	0.606530659712633\\
125	0.706648277857716\\
126	0.800737402916808\\
127	0.882496902584595\\
128	0.945959468906765\\
129	0.986207116743916\\
130	1\\
131	0.986207116743916\\
132	0.945959468906765\\
133	0.882496902584595\\
134	0.800737402916808\\
135	0.706648277857716\\
136	0.606530659712633\\
137	0.506335616648101\\
138	0.411112290507187\\
139	0.32465246735835\\
140	0.249352208777296\\
141	0.186270463697701\\
142	0.135335283236613\\
143	0.0956344448325386\\
144	0.0657285286165304\\
145	0.0439369336234074\\
146	0.0285655007845504\\
147	0.0180630134197813\\
148	0.0111089965382423\\
149	0.0066450112827414\\
150	0.0038659201394728\\
151	0.00218749111818289\\
152	0.0012038599948282\\
153	0.000644379820568608\\
154	0.000335462627902512\\
155	0.00016985667656141\\
156	8.36483472297277e-05\\
157	4.00652973929511e-05\\
158	1.86644691135205e-05\\
159	8.45666596870112e-06\\
160	3.72665317207867e-06\\
161	1.59725788283431e-06\\
162	6.65836146985732e-07\\
163	2.69957850336301e-07\\
164	1.0645371411076e-07\\
165	4.08283604114252e-08\\
166	1.52299797447126e-08\\
167	5.52551772036671e-09\\
168	1.94976778601724e-09\\
169	6.69158609129278e-10\\
170	2.23363143620316e-10\\
171	7.25153956932532e-11\\
172	2.28973484564555e-11\\
173	7.03196062084075e-12\\
174	2.10040929273226e-12\\
175	6.10193667760532e-13\\
176	1.72412093903551e-13\\
177	4.73809779628813e-14\\
178	1.26641655490942e-14\\
179	3.29219387117252e-15\\
180	8.32396967698111e-16\\
181	2.0469717131642e-16\\
182	4.89586526458611e-17\\
183	1.13889408651823e-17\\
184	2.57675710915498e-18\\
185	5.67021977492392e-19\\
186	1.21356366812579e-19\\
187	2.52616378092569e-20\\
188	5.11442373180375e-21\\
189	1.00708974473355e-21\\
190	1.92874984796392e-22\\
191	3.59269128537111e-23\\
192	6.50878851546785e-24\\
193	1.1468765822256e-24\\
194	1.96548382416367e-25\\
195	3.27611039093474e-26\\
196	5.3110922496791e-27\\
197	8.37424045460638e-28\\
198	1.2842313672676e-28\\
199	1.91547895196715e-29\\
200	2.77873901831711e-30\\
};
\addlegendentry{Analytic Solution}

\addplot [color=red, dashed, line width=0.7pt]
  table[row sep=crcr]{%
1	-1.91547895196715e-29\\
5	-3.27611039093474e-26\\
9	-3.59269128537111e-23\\
13	-2.52616378092569e-20\\
17	-1.13889408651823e-17\\
21	-3.29219387117252e-15\\
25	-6.10193667760532e-13\\
29	-7.25153956932532e-11\\
33	-5.52551772036671e-09\\
37	-2.69957850336301e-07\\
41	-8.45666596870112e-06\\
45	-0.00016985667656141\\
49	-0.00218749111818289\\
53	-0.0180630134197813\\
57	-0.0956344448325386\\
61	-0.32465246735835\\
65	-0.706648277857716\\
69	-0.986207116743916\\
73	-0.882496902584595\\
77	-0.506335616648101\\
81	-0.186270463697698\\
85	-0.0439369336227972\\
89	-0.00664501121022601\\
93	-0.000644374295050887\\
97	-3.97953395426148e-05\\
101	6.85940808586681e-06\\
105	0.000169815848200999\\
109	0.00218749044902428\\
113	0.0180630134127493\\
117	0.0956344448324913\\
121	0.32465246735835\\
125	0.706648277857716\\
129	0.986207116743916\\
133	0.882496902584595\\
137	0.506335616648101\\
141	0.186270463697701\\
145	0.0439369336234074\\
149	0.0066450112827414\\
153	0.000644379820568608\\
157	4.00652973929511e-05\\
161	1.59725788283431e-06\\
165	4.08283604114252e-08\\
169	6.69158609129278e-10\\
173	7.03196062084075e-12\\
177	4.73809779628813e-14\\
181	2.0469717131642e-16\\
185	5.67021977492392e-19\\
189	1.00708974473355e-21\\
193	1.1468765822256e-24\\
197	8.37424045460638e-28\\
};
\addlegendentry{FDTD Method}

\end{axis}
\end{tikzpicture}%
    \caption{Magnetic field.}
\end{figure}
Figs.~9 and 10 show the agreement between the FDTD and the analytic solution for the electromagnetic fields. A Gaussian pulse is generated at the middle of the free space medium and propagates outward in both directions. 
\begin{figure}[H]
\centering
\setlength\fheight{5cm}
\setlength\fwidth{8cm}
\input{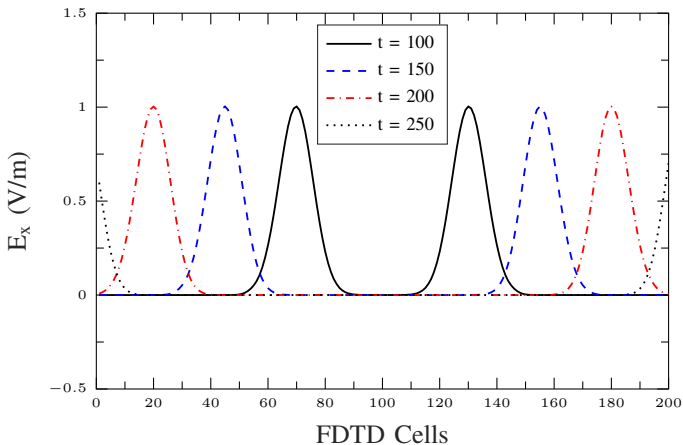}
    \caption{Electric field propagation.}
\end{figure}
\begin{figure}[H]
\centering
\setlength\fheight{5cm}
\setlength\fwidth{8cm}
\input{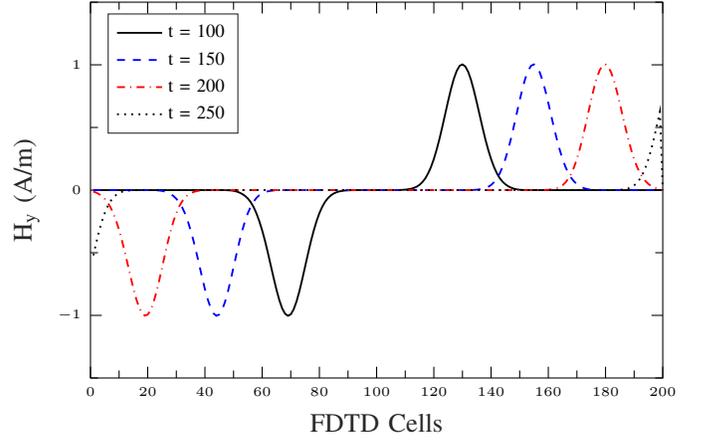}
    \caption{Magnetic field propagation.}
\end{figure}
Figs.~11 and 12 illustrate the propagation of the Gaussian pulse in free space medium bounded by absorbing boundaries in different instants. The wave starts from the middle and then it propagates in the two directions until the instant t = 250. At the instant t = 150, the wave moves to other positions and it is still far from the boundaries. At the instant t = 200, the wave starts hitting the two boundaries and it roughly overtakes the two ends at the instant t = 250 and then it totally vanishes from the space problem because it is absorbed by the two boundaries.


\section{Two-Dimentional Simulation}
\subsection{Free Space Medium with Absorbing Boundaries}
 We will start with the normalized Maxwell's equations as follows
 \begin{align}
 \frac{\partial \widetilde{\textbf{\emph{D}}}}{\partial t} = \frac{1}{\sqrt{\epsilon_0 \mu_0}} \nabla \times \widetilde{\textbf{\emph{H}}},
 \end{align}
 \begin{align}
 \widetilde{\textbf{\emph{D}}}(\omega) = \bold{\epsilon}^*_r(\omega) \cdot  \widetilde{\textbf{\emph{E}}}(\omega),
 \end{align}
 \begin{align}
  \frac{\partial \textbf{\emph{H}}}{\partial t} = - \frac{1}{\sqrt{\epsilon_0 \mu_0}} \nabla \times \widetilde{\textbf{\emph{E}}},
 \end{align}
 For the TM mode, the above equations reduced as follows
 \begin{align}
 \frac{\partial D_z}{\partial t} = \frac{1}{\sqrt{\epsilon_0 \mu_0}}\left(\frac{\partial H_y}{\partial x} - \frac{\partial H_x}{\partial y}  \right),
 \end{align}
 \begin{align}
 \textbf{\emph{D}}(\omega) = \bold{\epsilon}^*_r(\omega) \cdot \textbf{\emph{E}}(\omega),
 \end{align}
 \begin{align}
 \frac{\partial H_x}{\partial t} = -\frac{1}{\sqrt{\epsilon_0 \mu_0}} \frac{\partial E_z}{\partial y},
 \end{align}
 \begin{align}
 \frac{\partial H_y}{\partial t} = \frac{1}{\sqrt{\epsilon_0 \mu_0}} \frac{\partial E_z}{\partial x},
 \end{align}
The discretization of the above equations leads to
\begin{align}
\frac{D_z^{n+1/2}(i,j) - D_z^{n-1/2}(i,j)}{\Delta t} = \ddfrac{1}{\sqrt{\epsilon_0 \mu_0}}\\ \times~ \left( \frac{H_x^n\left(i+ \frac{1}{2},j\right) - H_y^n\left(i - \frac{1}{2},j \right)}{\Delta x}  \right) - \frac{1}{\sqrt{\epsilon_0 \mu_0}}\\ \times~ \left( \frac{H_x^n\left(i ,j+ \frac{1}{2}\right) - H_x^n\left(i ,j- \frac{1}{2} \right)}{\Delta x}  \right), 
\end{align}
\begin{align}
\frac{H_x^{n+1}\left(i +\frac{1}{2},j \right) - H_x^n\left(i + \frac{1}{2},j\right)}{\Delta t} = -\frac{1}{\sqrt{\epsilon_0 \mu_0}}\\ \times~\frac{E_z^{n+1/2}(i,j+1) - E_z^{n+1/2}(i,j)}{\Delta x},
\end{align}

\begin{align}
\frac{H_y^{n+1}\left(i +\frac{1}{2},j \right) - H_y^n\left(i + \frac{1}{2},j\right)}{\Delta t} = -\frac{1}{\sqrt{\epsilon_0 \mu_0}}\\ \times~\frac{E_z^{n+1/2}(i+1,j) - E_z^{n+1/2}(i,j)}{\Delta x},
\end{align}

\begin{figure*}
\begin{center}
\begin{subfigure}[b]{.3\linewidth}
  \centering
  \setlength\fheight{0.6\textwidth}
  \setlength\fwidth{1.0\textwidth}
  \input{fr1.tikz}
  \caption{T = 20.}
  \label{fig:mse_omegadot}
\end{subfigure}\hfill
\begin{subfigure}[b]{.3\linewidth}
  \centering
  \setlength\fheight{0.6\textwidth}
  \setlength\fwidth{1.0\textwidth}
  \input{fr2.tikz}
  \caption{T = 30.}
  \label{fig:mse_p}
\end{subfigure}\hfill
\begin{subfigure}[b]{.3\linewidth}
  \centering
  \setlength\fheight{0.6\textwidth}
  \setlength\fwidth{1.0\textwidth}
  \input{fr3.tikz}
  \caption{T = 40.}
  \label{fig:mse_p}
\end{subfigure}\\
\begin{subfigure}[b]{.3\linewidth}
  \centering
  \setlength\fheight{0.6\textwidth}
  \setlength\fwidth{1.0\textwidth}
  \input{fr4.tikz}
  \caption{T = 50.}
  \label{fig:mse_q}
\end{subfigure}\hfill
\begin{subfigure}[b]{.3\linewidth}
  \centering
  \setlength\fheight{0.6\textwidth}
  \setlength\fwidth{1.0\textwidth}
  \input{fr5.tikz}
  \caption{T = 80.}
  \label{fig:mse_alpha}
\end{subfigure}\hfill
\begin{subfigure}[b]{.3\linewidth}
  \centering
  \setlength\fheight{0.6\textwidth}
  \setlength\fwidth{1.0\textwidth}
  \input{fr6.tikz}
  \caption{T = 100.}
  \label{fig:mse_alpha}
\end{subfigure}
\end{center}
\caption{FDTD-2D simulation of a Gaussian pulse in free space medium with absorbing boundaries in various time instants.}
\end{figure*}
Fig.~13 shows the propagation of a Gaussian pulse generated in the middle of the space problem. We observe how the wave propagates in all directions till it hits the boundaries and then totally vanishes. 
\subsection{The Perfectly Matched Layer (PML)}
The PML parameters can be presented as follows
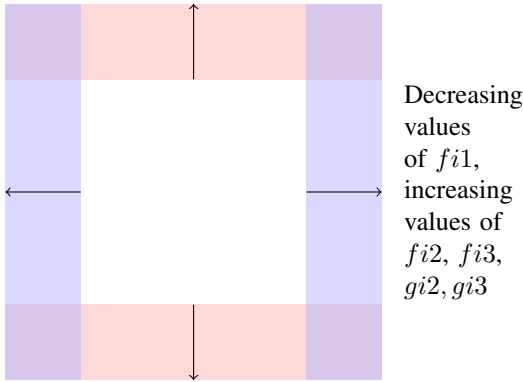
\begin{figure}[H]
\centering
\setlength\fheight{5cm}
\setlength\fwidth{8cm}
\begin{tikzpicture}
  
  \fill [red!30!white,opacity=0.5] (0,0) rectangle (5,1);
  \fill [red!30!white,opacity=0.5] (0,4) rectangle (5,5);
  \fill [blue!30!white,opacity=0.5] (0,0) rectangle (1,5);
  \fill [blue!30!white,opacity=0.5] (4,0) rectangle (5,5);
  
  \draw[->] (2.5,4) -- (2.5,5);
  \draw[->] (2.5,1) -- (2.5,0);
  \draw[->] (1,2.5) -- (0,2.5);
  \draw[->] (4,2.5) -- (5,2.5);
  \node[text width=3cm] at (0,5.7) { The corners are an overlap of both sets of parameters};  
  
  \node[text width=4cm] at (5.5,5.7) { Decreasing values of $fj1$, increasing values of $fj2, fj3, gj2, gj3$};
  
  \node[text width=2cm] at (6.3,2.5) { Decreasing values of $fi1$, increasing values of $fi2$, $fi3$, $gi2, gi3$};  
\end{tikzpicture}
    \caption{Parameters related to PML.}
\end{figure}
Maxwell's equations can be updated as
\begin{align}
\begin{split}
D_z^{n+1/2}(i,j) =& gi3(i)gj3(j)D_z^{n-1/2}(i,j) + 0.5gi2(i)gj2(j)\\& \times~\left[ H_y^n\left(i+\frac{1}{2},j\right) - H_y^n\left(i-\frac{1}{2},j\right) \right. \\& \left. - H_y^n\left(i,j+\frac{1}{2}\right) + H_y^n\left(i,j-\frac{1}{2}\right) \right], 
\end{split}
\end{align}
where $gi2, gi3, gj2$, and $gj3$ are given by
\begin{align}
gi2 = gj2 = \frac{1}{1 + \frac{\sigma \Delta t}{2 \epsilon_0}},
\end{align}
\begin{align}
gi3 = gj3 = \frac{1 - \frac{\sigma \Delta t}{2 \epsilon_0}}{1 + \frac{\sigma \Delta t}{2 \epsilon_0}},
\end{align}
The curl of the electric field can be given by
\begin{align}
\nabla \times \bold{E} = E_z^{n+1/2}(i+1,j) - E_z^{n+1/2}(i,j) ,    
\end{align}
The incident magnetic field in the $Y$-direction can be given by
\begin{align}
I^{n+1/2}_{H_y}\left(i+\frac{1}{2},j\right) = I^{n-1/2}_{H_y}\left(i+\frac{1}{2},j\right) + \nabla \times \bold{E},
\end{align}
The total magnetic field in the Y-direction can be evaluated recursively as follows
\begin{align}
\begin{split}
H_y^{n+1/2}\left(i+\frac{1}{2},j\right) = gi3\left(i+\frac{1}{2}\right) H_y^{n}\left(i+\frac{1}{2},j\right)\\ - 0.5 gi2\left(i+\frac{1}{2}\right)\nabla \times \bold{E} - gj1(j) I^{n+1/2}_{H_y}\left(i+\frac{1}{2},j\right), 
\end{split}
\end{align}
where $gi1$ is given by
\begin{align}
gi1 = \frac{\sigma \Delta t}{2\epsilon_0},
\end{align}
The incident magnetic field in the $X$-direction is given by
\begin{align}
I^{n+1/2}_{H_x}\left(i,j+\frac{1}{2}\right) = I^{n-1/2}_{H_x}\left(i,j+\frac{1}{2}\right) + \nabla \times \bold{E},
\end{align}
The total magnetic field in the $X$-direction can be obtained by
\begin{align}
\begin{split}
H_x^{n+1}\left(i,j+\frac{1}{2}\right) =& gj3\left(j+\frac{1}{2}\right) H_x^{n}\left(i,j+\frac{1}{2}\right)\\& +~0.5~ gj2\left(j+\frac{1}{2}\right)\nabla \times \bold{E}\\& +~ gj1(j) I^{n+1/2}_{H_y}\left(i,j+\frac{1}{2}\right),
\end{split}
\end{align}
\begin{figure*}
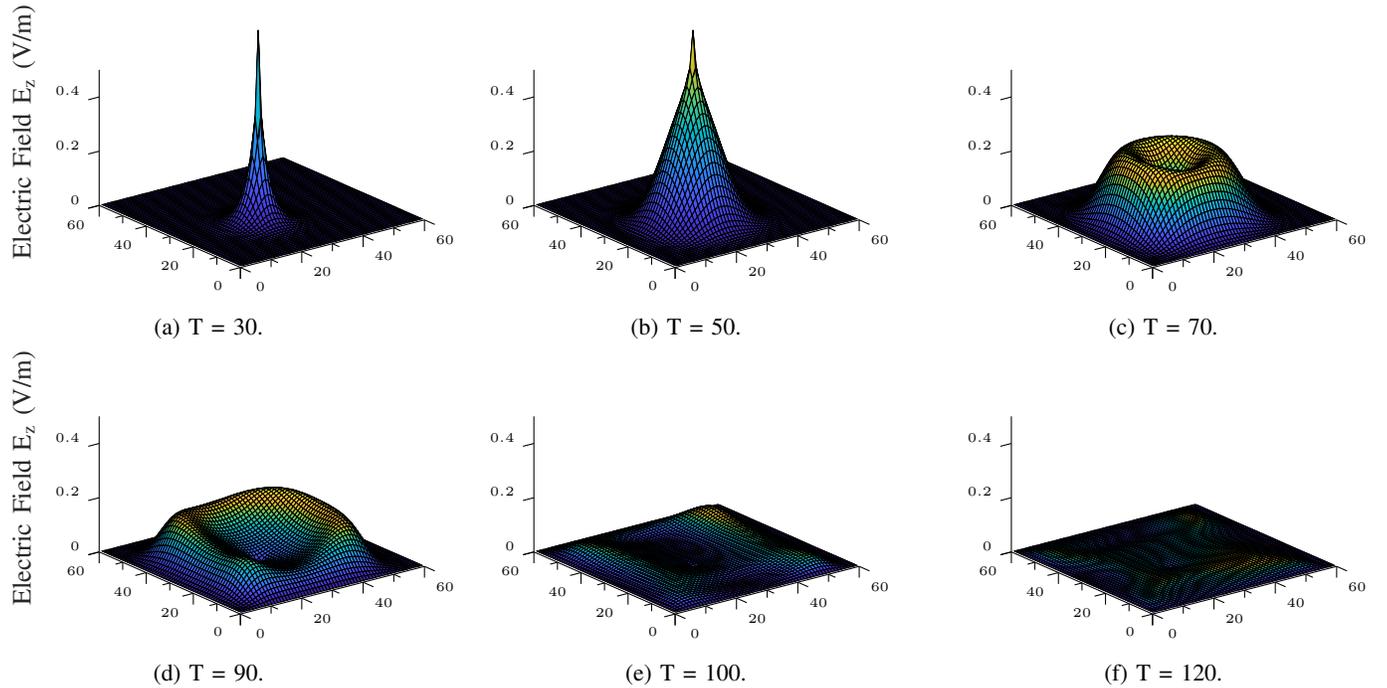

\begin{center}
\begin{subfigure}[b]{.3\linewidth}
  \centering
  \setlength\fheight{0.6\textwidth}
  \setlength\fwidth{1.0\textwidth}
  \input{pml1.tikz}
  \caption{T = 30.}
  \label{fig:mse_omegadot}
\end{subfigure}\hfill
\begin{subfigure}[b]{.3\linewidth}
  \centering
  \setlength\fheight{0.6\textwidth}
  \setlength\fwidth{1.0\textwidth}
  \input{pml2.tikz}
  \caption{T = 50.}
  \label{fig:mse_p}
\end{subfigure}\hfill
\begin{subfigure}[b]{.3\linewidth}
  \centering
  \setlength\fheight{0.6\textwidth}
  \setlength\fwidth{1.0\textwidth}
  \input{pml3.tikz}
  \caption{T = 70.}
  \label{fig:mse_p}
\end{subfigure}\\
\begin{subfigure}[b]{.3\linewidth}
  \centering
  \setlength\fheight{0.6\textwidth}
  \setlength\fwidth{1.0\textwidth}
  \input{pml4.tikz}
  \caption{T = 90.}
  \label{fig:mse_q}
\end{subfigure}\hfill
\begin{subfigure}[b]{.3\linewidth}
  \centering
  \setlength\fheight{0.6\textwidth}
  \setlength\fwidth{1.0\textwidth}
  \input{pml5.tikz}
  \caption{T = 100.}
  \label{fig:mse_alpha}
\end{subfigure}\hfill
\begin{subfigure}[b]{.3\linewidth}
  \centering
  \setlength\fheight{0.6\textwidth}
  \setlength\fwidth{1.0\textwidth}
  \input{pml6.tikz}
  \caption{T = 120.}
  \label{fig:mse_alpha}
\end{subfigure}
\end{center}
\caption{FDTD-2D simulation of a Gaussian pulse with PML in various instants.}
\end{figure*}
Fig.~15 shows the FDTD-2D simulation of a Gaussian source initiated at a point that is offset five cells from the center of the problem space in each direction. As the wave reaches the PML, which is eight cells on every side, it is absorbed. The effectiveness of the PML is apparent in the bottom figure because the contours would not be concentric circles if the outgoing wave was partially reflected.

\subsection{Total/Scattered Field Formulation}
In order to simulate a plane wave in a two-dimensional FDTD program, the problem space will be divided into two regions, the total field and the scattered field in Fig.~16. The two primary reasons for doing this are (i) the propagating plane wave should not interact with the ABCs and (ii) the load on the ABCs should be minimized. These boundary conditions are not perfect, that is, a certain portion of the impinging wave is reflected back into the problem space. By subtracting the incident field, the amount of the radiating field hitting the boundary is minimized, thereby reducing the amount of error.
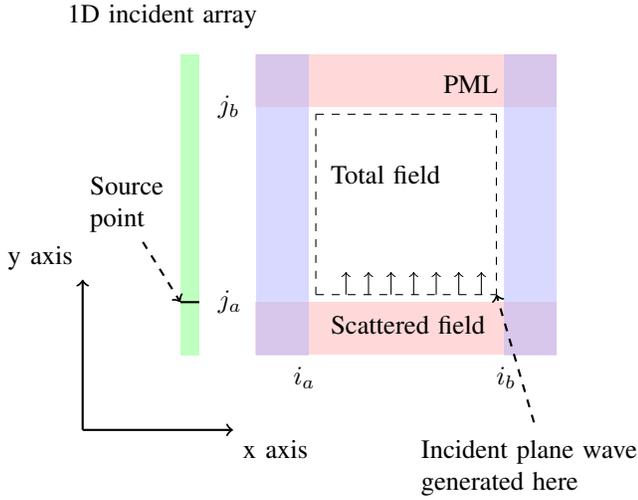
\begin{figure}[H]
\centering
\setlength\fheight{5cm}
\setlength\fwidth{8cm}
\begin{tikzpicture}
  
  \fill [red!30!white,opacity=0.5] (0,0) rectangle (4,.7);
  \fill [red!30!white,opacity=0.5] (0,3.3) rectangle (4,4);
  \fill [blue!30!white,opacity=0.5] (0,0) rectangle (0.7,4);
  \fill [blue!30!white,opacity=0.5] (3.3,0) rectangle (4,4);
  
  \fill [green!50!white,opacity=0.5] (-1,0) rectangle (-0.75,4);
  
  \node[text width=3cm] at (2,-0.3) { $i_a$}; 
  \node[text width=3cm] at (4.7,-0.3) { $i_b$};
  \node[text width=3cm] at (1,0.7) { $j_a$}; 
  \node[text width=3cm] at (1,3.3) { $j_b$};
  \node[text width=3cm] at (2.5,.4) { Scattered field};
  \node[text width=3cm] at (2.5,2.4) { Total field};
  \node[text width=3cm] at (4,3.6) { PML};

  

\draw[dashed] (.8,.8) -- (3.2,.8);
\draw[dashed] (.8,.8) -- (.8,3.2);
\draw[dashed] (.8,3.2) -- (3.2,3.2);
\draw[dashed] (3.2,3.2) -- (3.2,.8);

\draw[->] (1.2,.8) -- (1.2,1.1);
\draw[->] (1.5,.8) -- (1.5,1.1);
\draw[->] (1.8,.8) -- (1.8,1.1);
\draw[->] (2.1,.8) -- (2.1,1.1);
\draw[->] (2.4,.8) -- (2.4,1.1);
\draw[->] (2.7,.8) -- (2.7,1.1);
\draw[->] (3,.8) -- (3,1.1);

\draw[thick,solid] (-1,.7) -- (-.75,.7);
  
\draw[thick,dashed,->] (3.7,-0.9) -- (3.2,0.8);
\node[text width=3cm] at (3.7,-1.5) { Incident plane wave generated here}; 
 
\node[text width=3cm] at (-1,4.5) { 1D incident array};
\node[text width=1cm] at (-1.7,2) { Source point};
\draw[thick,dashed,->] (-1.5,1.5) -- (-1,.7);  
    
\draw[thick,->] (-2.3,-1) -- (-0.3,-1) node[anchor=north west] {x axis};
\draw[thick,->] (-2.3,-1) -- (-2.3,1) node[anchor=south east] {y axis};

\end{tikzpicture}
    \caption{Total field/scattered field of the two-dimensional space.}
\end{figure}
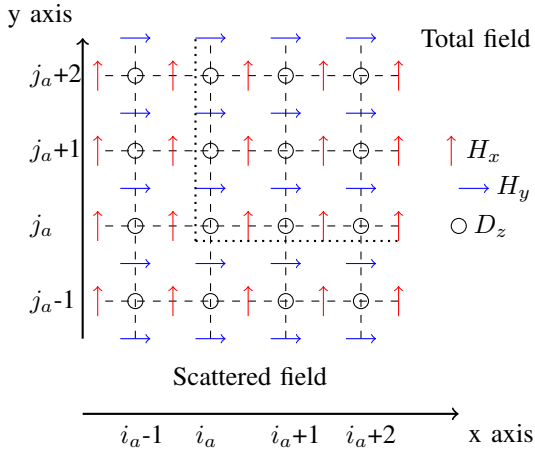
\begin{figure}[H]
\centering
\setlength\fheight{5cm}
\setlength\fwidth{8cm}
\begin{tikzpicture}

\draw[thick,->] (0,0) -- (5,0) node[anchor=north west] {x axis};
\draw[thick,->] (0,1) -- (0,5) node[anchor=south east] {y axis};

\node[text width=3cm] at (0.8,1.5) { $j_a$-1}; 
\node[text width=3cm] at (0.8,2.5) { $j_a$};  
\node[text width=3cm] at (0.8,3.5) { $j_a$+1};
\node[text width=3cm] at (0.8,4.5) { $j_a$+2};  

\node[text width=3cm] at (2,-.3) { $i_a$-1}; 
\node[text width=3cm] at (3,-.3) { $i_a$}; 
\node[text width=3cm] at (4,-.3) { $i_a$+1}; 
\node[text width=3cm] at (5,-.3) { $i_a$+2}; 

\node[text width=3cm] at (2.7,.5) { Scattered field};  

\draw[dashed] (0.3,1.5) -- (4.3,1.5);
\draw[dashed] (0.3,2.5) -- (4.3,2.5);
\draw[dashed] (0.3,3.5) -- (4.3,3.5);
\draw[dashed] (0.3,4.5) -- (4.3,4.5);

\draw[dashed] (.7,1) -- (.7,5);
\draw[dashed] (1.7,1) -- (1.7,5);
\draw[dashed] (2.7,1) -- (2.7,5);
\draw[dashed] (3.7,1) -- (3.7,5);

\draw (.7,1.5) circle (0.1cm);
\draw (.7,2.5) circle (0.1cm);
\draw (.7,3.5) circle (0.1cm);
\draw (.7,4.5) circle (0.1cm);

\draw (1.7,1.5) circle (0.1cm);
\draw (2.7,1.5) circle (0.1cm);
\draw (3.7,1.5) circle (0.1cm);

\draw (1.7,2.5) circle (0.1cm);
\draw (1.7,3.5) circle (0.1cm);
\draw (1.7,4.5) circle (0.1cm);

\draw (2.7,2.5) circle (0.1cm);
\draw (2.7,3.5) circle (0.1cm);
\draw (2.7,4.5) circle (0.1cm);

\draw (3.7,2.5) circle (0.1cm);
\draw (3.7,3.5) circle (0.1cm);
\draw (3.7,4.5) circle (0.1cm);

\draw[->,blue] (.5,1) -- (.9,1);
\draw[->,blue] (.5,2) -- (.9,2);
\draw[->,blue] (.5,3) -- (.9,3);
\draw[->,blue] (.5,4) -- (.9,4);
\draw[->,blue] (.5,5) -- (.9,5);

\draw[->,blue] (1.5,1) -- (1.9,1);
\draw[->,blue] (1.5,2) -- (1.9,2);
\draw[->,blue] (1.5,3) -- (1.9,3);
\draw[->,blue] (1.5,4) -- (1.9,4);
\draw[->,blue] (1.5,5) -- (1.9,5);

\draw[->,blue] (2.5,1) -- (2.9,1);
\draw[->,blue] (2.5,2) -- (2.9,2);
\draw[->,blue] (2.5,3) -- (2.9,3);
\draw[->,blue] (2.5,4) -- (2.9,4);
\draw[->,blue] (2.5,5) -- (2.9,5);

\draw[->,blue] (3.5,1) -- (3.9,1);
\draw[->,blue] (3.5,2) -- (3.9,2);
\draw[->,blue] (3.5,3) -- (3.9,3);
\draw[->,blue] (3.5,4) -- (3.9,4);
\draw[->,blue] (3.5,5) -- (3.9,5);

\draw[->,blue] (5,3) -- (5.4,3);
\node[text width=3cm] at (7,3) { $H_y$};  

\draw[->,red] (.2,1.3) -- (.2,1.7);
\draw[->,red] (1.2,1.3) -- (1.2,1.7);
\draw[->,red] (2.2,1.3) -- (2.2,1.7);
\draw[->,red] (3.2,1.3) -- (3.2,1.7);
\draw[->,red] (4.2,1.3) -- (4.2,1.7);

\draw[->,red] (.2,2.3) -- (.2,2.7);
\draw[->,red] (.2,3.3) -- (.2,3.7);
\draw[->,red] (.2,4.3) -- (.2,4.7);

\draw[->,red] (1.2,2.3) -- (1.2,2.7);
\draw[->,red] (1.2,3.3) -- (1.2,3.7);
\draw[->,red] (1.2,4.3) -- (1.2,4.7);

\draw[->,red] (2.2,2.3) -- (2.2,2.7);
\draw[->,red] (2.2,3.3) -- (2.2,3.7);
\draw[->,red] (2.2,4.3) -- (2.2,4.7);

\draw[->,red] (3.2,2.3) -- (3.2,2.7);
\draw[->,red] (3.2,3.3) -- (3.2,3.7);
\draw[->,red] (3.2,4.3) -- (3.2,4.7);

\draw[->,red] (4.2,2.3) -- (4.2,2.7);
\draw[->,red] (4.2,3.3) -- (4.2,3.7);
\draw[->,red] (4.2,4.3) -- (4.2,4.7);

\draw[->,red] (4.9,3.3) -- (4.9,3.7);
\node[text width=3cm] at (6.6,3.5) { $H_x$}; 

\draw (5,2.5) circle (0.1cm);
\node[text width=3cm] at (6.7,2.5) { $D_z$};

\node[text width=3cm] at (6,5) { Total field}; 

\draw[dotted,thick] (1.5,2.3) -- (1.5,5);
\draw[dotted,thick] (1.5,2.3) -- (4.2,2.3);
  
\end{tikzpicture}
    \caption{Every point is in either the total field or the scattered field.}
\end{figure}
There are three places that must be modified\\
The $D_z$ value at $j = j_a$ or $j = j_b$
\begin{align}
D_z(i,j_a) = D_z(i,j_a) + 0.5 H_{x,inc}\left(j_a - \frac{1}{2} \right),    
\end{align}
\begin{align}
D_z(i,j_b) = D_z(i,j_b) - 0.5 H_{x,inc}\left(j_a - \frac{1}{2} \right),    
\end{align}
The $H_x$ field just outside at $j = j_a$ or $j = j_b$
\begin{align}
H_x\left(i,j_a -\frac{1}{2}\right) = H_x\left(i,j_a -\frac{1}{2}\right) + 0.5 E_{z,inc}(j_a),
\end{align}
\begin{align}
H_x\left(i,j_b +\frac{1}{2}\right) = H_x\left(i,j_b +\frac{1}{2}\right) - 0.5 E_{z,inc}(j_b),
\end{align}
$H_y$ just outside at $i = i_a$ or $i = i_b$
\begin{align}
H_y\left(i_a-\frac{1}{2},j\right) = H_y\left(i_a-\frac{1}{2},j\right) - 0.5 E_{z,inc}(j) ,
\end{align}
\begin{align}
H_y\left(i_b-\frac{1}{2},j\right) = H_y\left(i_b-\frac{1}{2},j\right) + 0.5 E_{z,inc}(j), 
\end{align}
\begin{figure*}
\begin{center}
\begin{subfigure}[b]{.3\linewidth}
  \centering
  \setlength\fheight{0.6\textwidth}
  \setlength\fwidth{1.0\textwidth}
  \input{plane1.tikz}
  \caption{T = 20.}
  \label{fig:mse_omegadot}
\end{subfigure}\hfill
\begin{subfigure}[b]{.3\linewidth}
  \centering
  \setlength\fheight{0.6\textwidth}
  \setlength\fwidth{1.0\textwidth}
  \input{plane2.tikz}
  \caption{T = 50.}
  \label{fig:mse_p}
\end{subfigure}\hfill
\begin{subfigure}[b]{.3\linewidth}
  \centering
  \setlength\fheight{0.6\textwidth}
  \setlength\fwidth{1.0\textwidth}
  \input{plane3.tikz}
  \caption{T = 90.}
  \label{fig:mse_p}
\end{subfigure}\\
\begin{subfigure}[b]{.3\linewidth}
  \centering
  \setlength\fheight{0.6\textwidth}
  \setlength\fwidth{1.0\textwidth}
  \input{plane4.tikz}
  \caption{T = 120.}
  \label{fig:mse_q}
\end{subfigure}\hfill
\begin{subfigure}[b]{.3\linewidth}
  \centering
  \setlength\fheight{0.6\textwidth}
  \setlength\fwidth{1.0\textwidth}
  \input{plane5.tikz}
  \caption{T = 150.}
  \label{fig:mse_alpha}
\end{subfigure}\hfill
\begin{subfigure}[b]{.3\linewidth}
  \centering
  \setlength\fheight{0.6\textwidth}
  \setlength\fwidth{1.0\textwidth}
  \input{plane6.tikz}
  \caption{T = 200.}
  \label{fig:mse_alpha}
\end{subfigure}
\end{center}
\caption{Simulation of a plane wave pulse propagating in free space. The incident pulse is generated at one end and subtracted at the other end.}
\end{figure*}
The simulation results are shown by Fig.~18.
\subsection{A Plane Wave Impinging on a Dielectric Cylinder}
To simulate a plane wave interacting with an object, we must specify the object according to its electromagnetic properties—the dielectric constant and the conductivity. For instance, suppose we are simulating a plane wave striking a dielectric cylinder 20 cm in diameter, which has a dielectric constant specified by the parameter epsilon and a conductivity specified by the parameter sigma.
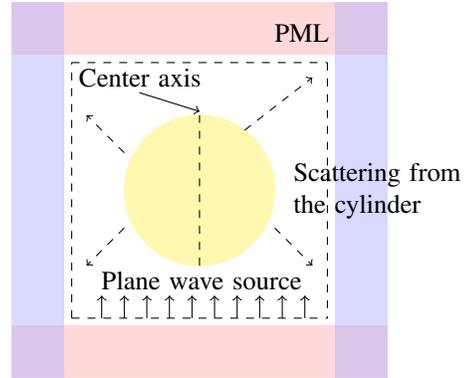
\begin{figure}[H]
\centering
\setlength\fheight{5cm}
\setlength\fwidth{8cm}
\begin{tikzpicture}
  
  \fill [red!30!white,opacity=0.5] (0,0) rectangle (5,.7);
  \fill [red!30!white,opacity=0.5] (0,4.3) rectangle (5,5);
  \fill [blue!30!white,opacity=0.5] (0,0) rectangle (0.7,5);
  \fill [blue!30!white,opacity=0.5] (4.3,0) rectangle (5,5);

  \node[text width=3cm] at (2.7,1.3) {Plane wave source};
  \node[text width=3cm] at (5,4.6) { PML};
  \node[text width=3cm] at (2.4,4) { Center axis};
  \draw[->] (1.7,3.8) -- (2.5,3.55);

  

\draw[dashed] (.8,.8) -- (4.2,.8);
\draw[dashed] (.8,.8) -- (.8,4.2);
\draw[dashed] (.8,4.2) -- (4.2,4.2);
\draw[dashed] (4.2,4.2) -- (4.2,.8);

\draw[->] (1.2,.8) -- (1.2,1.1);
\draw[->] (1.5,.8) -- (1.5,1.1);
\draw[->] (1.8,.8) -- (1.8,1.1);
\draw[->] (2.1,.8) -- (2.1,1.1);
\draw[->] (2.4,.8) -- (2.4,1.1);
\draw[->] (2.7,.8) -- (2.7,1.1);
\draw[->] (3,.8) -- (3,1.1);
\draw[->] (3.3,.8) -- (3.3,1.1);
\draw[->] (3.6,.8) -- (3.6,1.1);
\draw[->] (3.9,.8) -- (3.9,1.1);

\draw [fill ,yellow!40](2.5,2.5) circle (1cm);

\draw[dashed] (2.5,1.5) -- (2.5,3.5);

\draw[->,dashed] (3.1,3.3) -- (4,4);
\draw[->,dashed] (3.5,2) -- (4,1.5);  
\draw[->,dashed] (1.5,2) -- (1,1.5);
\draw[->,dashed] (1.5,3) -- (1,3.5);
\node[text width=2.5cm] at (5,2.5) {Scattering from the cylinder};

\end{tikzpicture}
    \caption{Simulation of a plane wave striking a dielectric cylinder. The fields scattered from the cylinder are the only fields to leave the total field and strike the PML.}
\end{figure}
The simulation of a plane wave pulse hitting a dielectric cylinder with $\epsilon_r$ = 30 and $\sigma$ = 0.3 is shown in Fig. 20. After 20 time steps, the plane wave has started from the side; after 50 time steps, the pulse is interacting with the cylinder. Some of the pulse passes through the cylinder, and some of it goes around it. After 150 steps, the main part of the propagating pulse is subtracted from the end of the total field.
\begin{figure*}
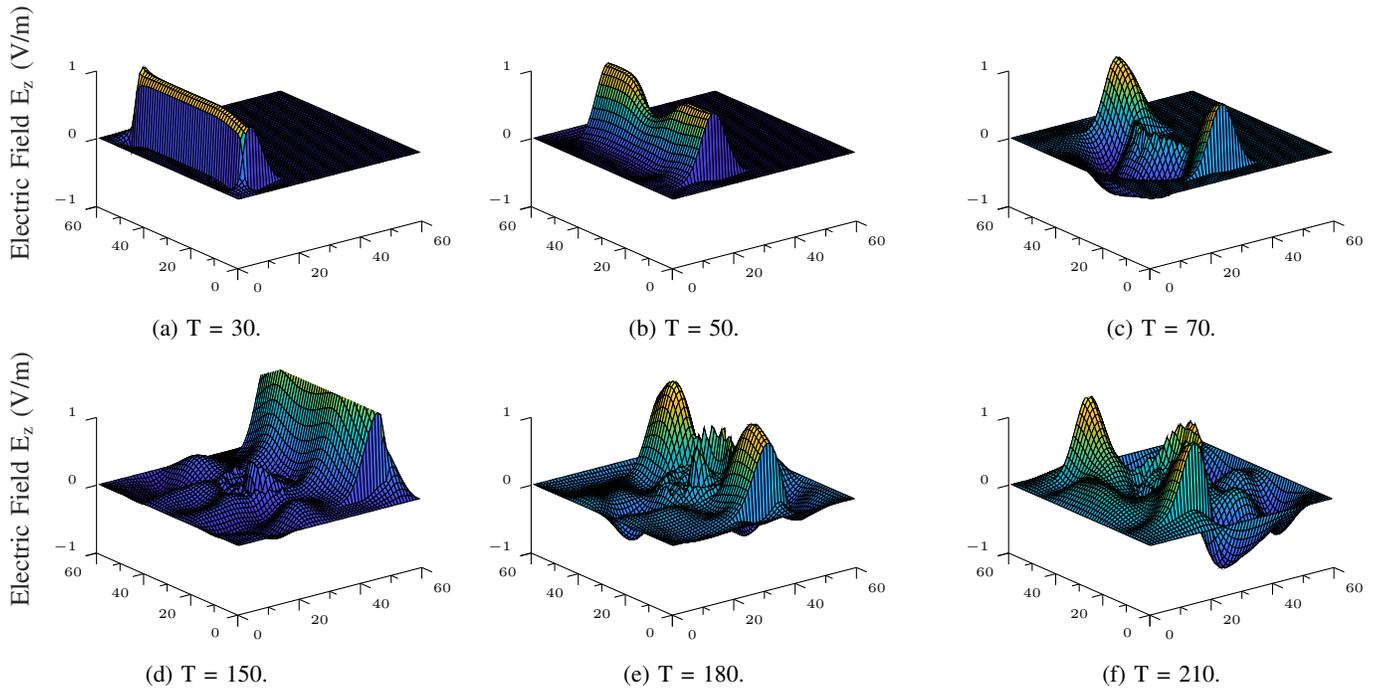

\begin{center}
\begin{subfigure}[b]{.3\linewidth}
  \centering
  \setlength\fheight{0.6\textwidth}
  \setlength\fwidth{1.0\textwidth}
  \input{cyl1.tikz}
  \caption{T = 30.}
  \label{fig:mse_omegadot}
\end{subfigure}\hfill
\begin{subfigure}[b]{.3\linewidth}
  \centering
  \setlength\fheight{0.6\textwidth}
  \setlength\fwidth{1.0\textwidth}
  \input{cyl2.tikz}
  \caption{T = 50.}
  \label{fig:mse_p}
\end{subfigure}\hfill
\begin{subfigure}[b]{.3\linewidth}
  \centering
  \setlength\fheight{0.6\textwidth}
  \setlength\fwidth{1.0\textwidth}
  \input{cyl3.tikz}
  \caption{T = 70.}
  \label{fig:mse_p}
\end{subfigure}\\
\begin{subfigure}[b]{.3\linewidth}
  \centering
  \setlength\fheight{0.6\textwidth}
  \setlength\fwidth{1.0\textwidth}
  \input{cyl4.tikz}
  \caption{T = 150.}
  \label{fig:mse_q}
\end{subfigure}\hfill
\begin{subfigure}[b]{.3\linewidth}
  \centering
  \setlength\fheight{0.6\textwidth}
  \setlength\fwidth{1.0\textwidth}
  \input{cyl5.tikz}
  \caption{T = 180.}
  \label{fig:mse_alpha}
\end{subfigure}\hfill
\begin{subfigure}[b]{.3\linewidth}
  \centering
  \setlength\fheight{0.6\textwidth}
  \setlength\fwidth{1.0\textwidth}
  \input{cyl6.tikz}
  \caption{T = 210.}
  \label{fig:mse_alpha}
\end{subfigure}
\end{center}
\caption{Simulation of a plane wave impinging on a dielectric cylinder. The cylinder is 20 cm in diameter and has a relative dielectric constant of 30 and a conductivity of 0.3 S/m.}
\end{figure*}
\subsection{Analytic Solution}
For the case of an electromagnetic wave propagating in free space medium bounded by PML, the space propagator factor $\Phi$ now consists of $k_x$ and $k_y$ (2 directions). It can be given by
\begin{equation}
\Phi = e^{-i(k_x~x + k_y~y)},    
\end{equation}
After reproducing the analysis of FDTD-1D for both the $X$ and $Y$ directions, the analytic solution can be derived. Regarding the case of a free space medium equipped with a dielectric cylinder, the scattered field can be derived as follows \cite{20}
\begin{align}
\begin{split}
E_s^{TM}(\bold{r}) =& (E_i^0\cdot \hat{v_i}) \frac{k}{k_{i,\rho}} \sum_{m=-\infty}^{\infty}\frac{J_m(k_{i,\rho},a)}{H_m^{(2)}(k_{i,\rho}a)}(-j)^m\exp(jm\phi_i)\\& \times~N^{H}_{m,k_{i,z}}(\rho,\phi,z),
\end{split}
\end{align}
\begin{align}
\begin{split}
H_s^{TM}(\bold{r}) = -\frac{j}{\eta}\sum_{m=-\infty}^{\infty}\nu_m \frac{J_m(k_{i,\rho})}{H_m^{(2)}(k_{i,\rho})}M^{H}_{m,k_{i,z}}(\rho,\phi,z),
\end{split}
\end{align}
where $\nu_m$, $M^{H}_{m,k_{i,z}}$, $N^{H}_{m,k_{i,z}}$, and $\psi_{m,k_{i,z}}$ are given by
\begin{align}
\nu_m = -(E_i^0\cdot \hat{v_i})\frac{k}{k_{i,\rho}}(-j)^m\exp(jm\phi_i),
\end{align}
\begin{align}
M^{H}_{m,k_{i,z}} = \frac{1}{k} \nabla \times [\psi^H_{m,k_{i,z}}\bold{z}],
\end{align}
\begin{align}
N^{H}_{m,k_{i,z}} = \frac{1}{k} \nabla \times [M^{H}_{m,k_{i,z}}],
\end{align}
\begin{align}
\psi^H_{m,k_{i,z}}(\rho,\phi,z) =& [A H_m^{(1)}(k_r \rho) + B H_m^{(2)}(k_r \rho)]\\&\times~ \exp[-jm\phi - jk_zz],
\end{align}
Where $H_m^{(1)}(\cdot)$, and $H_m^{(2)}(\cdot)$ are the Hankel functions of the first and second kind, respectively.
\section{Three-Dimensional Simulation}
\subsection{Free Space Simulation}
Starting with Maxwell's equations, we have
\begin{align}
 \frac{\partial \widetilde{\textbf{\emph{D}}}}{\partial t} = \frac{1}{\sqrt{\epsilon_0 \mu_0}} \nabla \times \widetilde{\textbf{\emph{H}}},
 \end{align}
 \begin{align}
 \widetilde{\textbf{\emph{D}}}(\omega) = \bold{\epsilon}^*_r(\omega) \cdot  \widetilde{\textbf{\emph{E}}}(\omega),
 \end{align}
 \begin{align}
  \frac{\partial \textbf{\emph{H}}}{\partial t} = - \frac{1}{\sqrt{\epsilon_0 \mu_0}} \nabla \times \widetilde{\textbf{\emph{E}}},
 \end{align}
 After expanding the above equations, we get
 \begin{align}
 \frac{\partial D_x}{\partial t} = \frac{1}{\sqrt{\epsilon_0 \mu_0}}\left(  \frac{\partial H_z}{\partial y} -  \frac{\partial H_y}{\partial z} \right),
 \end{align}
 \begin{align}
\frac{\partial D_y}{\partial t} = \frac{1}{\sqrt{\epsilon_0 \mu_0}}\left(  \frac{\partial H_x}{\partial z} -  \frac{\partial H_z}{\partial x} \right),  
 \end{align}
 \begin{align}
 \frac{\partial D_z}{\partial t} = \frac{1}{\sqrt{\epsilon_0 \mu_0}}\left(  \frac{\partial H_y}{\partial x} -  \frac{\partial H_x}{\partial y} \right),
 \end{align}
 \begin{align}
 \frac{\partial H_x}{\partial t} = \frac{1}{\sqrt{\epsilon_0 \mu_0}}\left(  \frac{\partial E_y}{\partial z} -  \frac{\partial E_z}{\partial y} \right),
 \end{align}
  \begin{align}
 \frac{\partial H_y}{\partial t} = \frac{1}{\sqrt{\epsilon_0 \mu_0}}\left(  \frac{\partial E_z}{\partial x} -  \frac{\partial E_x}{\partial z} \right),
 \end{align}
  \begin{align}
 \frac{\partial H_z}{\partial t} = \frac{1}{\sqrt{\epsilon_0 \mu_0}}\left(  \frac{\partial E_x}{\partial y} -  \frac{\partial E_y}{\partial x} \right),
 \end{align}
After taking the finite-difference approximations on the equations (69) and (72), we have
\begin{align}
\begin{split}
&D_z^{n+1/2}\left(i,j,k+\frac{1}{2}\right) = D_z^{n+1/2}\left(i,j,k+\frac{1}{2}\right) + \frac{\Delta t}{\Delta x\sqrt{\epsilon_0 \mu_0}} \\& \times~ \left[H_y^n\left(i+\frac{1}{2},j,k+\frac{1}{2}\right) - H_y^n\left(i-\frac{1}{2},j,k+\frac{1}{2}\right) \right. \\&\left. - H_x^{n+1}\left(i+\frac{1}{2},j+\frac{1}{2},k\right) + H_x^{n+1}\left(i+\frac{1}{2},j-\frac{1}{2},k\right)\right],
\end{split}
\end{align}
\begin{align}
\begin{split}
H_z^{n}\left(i+\frac{1}{2},j+\frac{1}{2},k\right) = H_z^{n}\left(i+\frac{1}{2},j+\frac{1}{2},k\right) + \frac{\Delta t}{\Delta x\sqrt{\epsilon_0 \mu_0}}\\ \times~\left[E_x^{n+1/2}\left(i+\frac{1}{2},j+1,k\right) - E_x^{n+1/2}\left(i+\frac{1}{2},j,k\right) \right. \\\left.- E_y^{n+1/2}\left(i+1,j+\frac{1}{2},k\right) + E_y^{n+1/2}\left(i,j+\frac{1}{2},k\right)   \right],
\end{split}
\end{align}
Fig.~21 shows the propagation of the $E_z$ from the microstrip in the XY plane level with the gap of the patch. Of course, there is radiation in the Z direction as well. This illustrates a major problem in three-dimensional simulations: unless one has unusually good graphics, visualizing three dimensions can be difficult.
\begin{figure*}
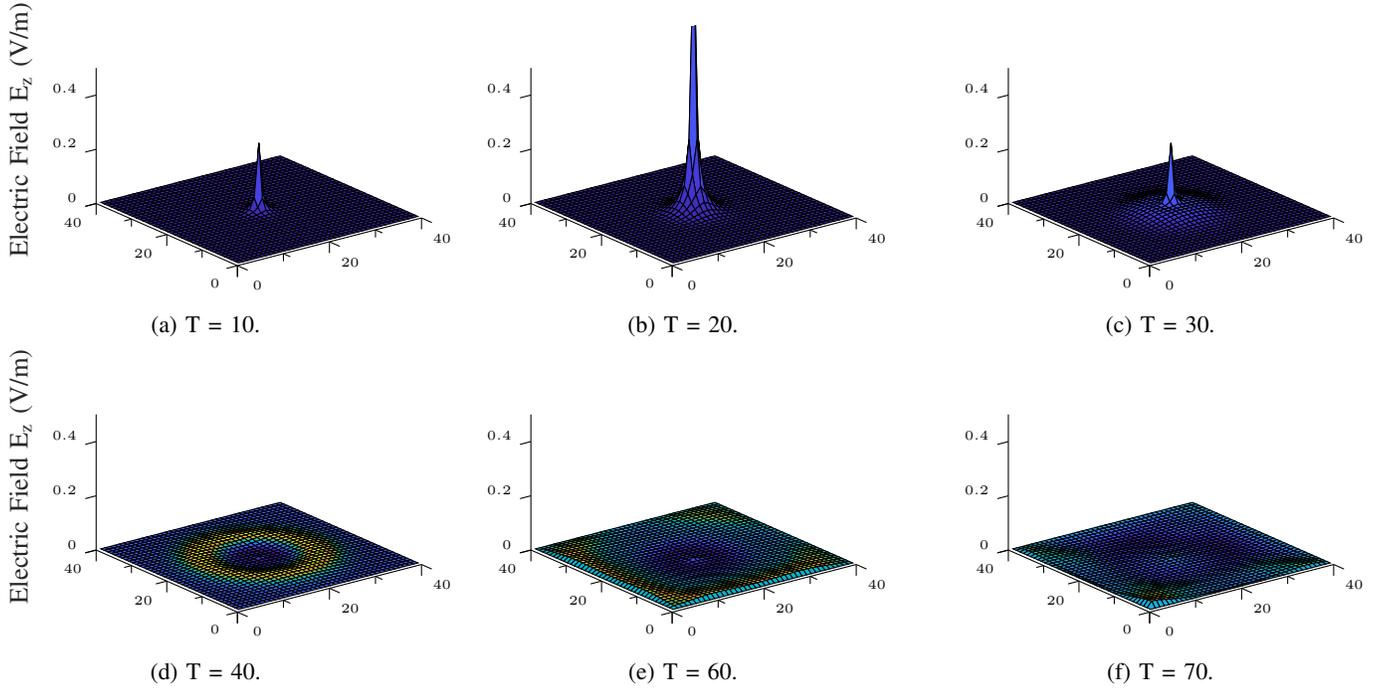

\begin{center}
\begin{subfigure}[b]{.3\linewidth}
  \centering
  \setlength\fheight{0.6\textwidth}
  \setlength\fwidth{1.0\textwidth}
  \input{3d1.tikz}
  \caption{T = 10.}
  \label{fig:mse_omegadot}
\end{subfigure}\hfill
\begin{subfigure}[b]{.3\linewidth}
  \centering
  \setlength\fheight{0.6\textwidth}
  \setlength\fwidth{1.0\textwidth}
  \input{3d2.tikz}
  \caption{T = 20.}
  \label{fig:mse_p}
\end{subfigure}\hfill
\begin{subfigure}[b]{.3\linewidth}
  \centering
  \setlength\fheight{0.6\textwidth}
  \setlength\fwidth{1.0\textwidth}
  \input{3d3.tikz}
  \caption{T = 30.}
  \label{fig:mse_p}
\end{subfigure}\\
\begin{subfigure}[b]{.3\linewidth}
  \centering
  \setlength\fheight{0.6\textwidth}
  \setlength\fwidth{1.0\textwidth}
  \input{3d4.tikz}
  \caption{T = 40.}
  \label{fig:mse_q}
\end{subfigure}\hfill
\begin{subfigure}[b]{.3\linewidth}
  \centering
  \setlength\fheight{0.6\textwidth}
  \setlength\fwidth{1.0\textwidth}
  \input{3d5.tikz}
  \caption{T = 60.}
  \label{fig:mse_alpha}
\end{subfigure}\hfill
\begin{subfigure}[b]{.3\linewidth}
  \centering
  \setlength\fheight{0.6\textwidth}
  \setlength\fwidth{1.0\textwidth}
  \input{3d6.tikz}
  \caption{T = 70.}
  \label{fig:mse_alpha}
\end{subfigure}
\end{center}
\caption{$E_z$ field radiation in a three-dimensional FDTD program.}
\end{figure*}
\subsection{A Plane Wave Impinging on a Dielectric Sphere in Free Space Medium Bounded by PML}
The Maxwell's equations in this case can be updated as follows
\begin{align}
\begin{split}
\nabla \times \bold{H} = H_y^n\left(i+\frac{1}{2},j,k+\frac{1}{2}\right) - H_y^n\left(i-\frac{1}{2},j,k+\frac{1}{2}\right) \\ - H_x^{n+1}\left(i+\frac{1}{2},j+\frac{1}{2},k\right) + H_x^{n+1}\left(i+\frac{1}{2},j-\frac{1}{2},k\right),    
\end{split}
\end{align}
The incident electric field density is given by
\begin{align}
I_{D_z}^{n}\left(i,j,k+\frac{1}{2} \right) = I_{D_z}^{n-1}\left(i,j,k+\frac{1}{2} \right) + \nabla \times \bold{H},
\end{align}
\begin{align}
\begin{split}
D_z^{n+1/2}\left(i,j,k+\frac{1}{2} \right) = gi3(i)gj3(j)D_z^{n+1/2}\left(i,j,k+\frac{1}{2} \right)\\ + 0.5gi2(i)gj2(j)\left[\nabla \times \bold{H} + gk1(k) I_{D_z}^{n+1/2}\left(i,j,k+\frac{1}{2} \right) \right],
\end{split}
\end{align}
Now that we have a program that generates a plane wave in three dimensions, we will want to start putting objects in the problem space to see how the plane wave interacts with them. In two dimensions, we chose a cylinder because we had an analytic solution with which we could check the accuracy of our calculation via a Bessel function expansion. It turns out that the interaction of a plane wave with a dielectric sphere can be determined by an expansion of the modified Bessel functions.
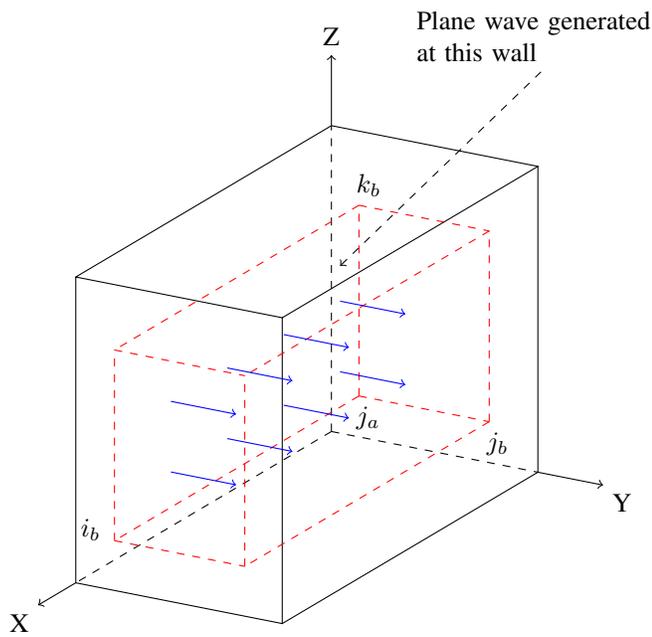
\begin{figure}[H]
\centering
\setlength\fheight{5cm}
\setlength\fwidth{8cm}
\tdplotsetmaincoords{70}{120}
\begin{tikzpicture}[tdplot_main_coords]
\def\BigSide{5}
\def\SmallSide{1.5}
\pgfmathsetmacro{\CalcSide}{\BigSide-\SmallSide}

\tdplotsetcoord{P}{sqrt(3)*\BigSide}{60}{25}

\coordinate (sxl) at (\BigSide,\CalcSide,\BigSide);
\coordinate (syl) at (\CalcSide,\CalcSide,\BigSide);
\coordinate (szl) at (\CalcSide,\BigSide,\BigSide);

\draw[dashed] 
  (0,0,0) -- (Px)
  (0,0,0) -- (Py)
  (0,0,0) -- (Pz);
\draw[->] 
  (Px) -- ++ (1,0,0) node[anchor=north east]{X};
\draw[->]
   (Py) -- ++(0,1,0) node[anchor=north west]{Y};
\draw[->] 
  (Pz) -- ++(0,0,1) node[anchor=south]{Z};

\draw
  (Pxz) -- (P) -- (Pxy) -- (Px) -- (Pxz) -- (Pz) -- (Pyz) -- (P); 
\draw
  (Pyz) -- (Py) -- (Pxy);

 \draw[dashed,red] (1,1,1) -- (7.5,1,1); 
 \draw[dashed,red](1,1,1) -- (1,3,1); 
 \draw[dashed,red](1,1,1) -- (1,1,3.7); 
 
 \draw[dashed,red] (1,3,1) -- (7.5,3,1); 
 \draw[dashed,red](7.5,1,1) -- (7.5,3,1); 
 \draw[dashed,red](1,3,1) -- (1,3,3.7); 
 
 \draw[dashed,red] (1,3,3.7) -- (7.5,3,3.7); 
 \draw[dashed,red](1,1,3.7) -- (7.5,1,3.7); 
 
 \draw[dashed,red](7.5,1,3.7) -- (7.5,3,3.7); 
 \draw[dashed,red](1,1,3.7) -- (1,3,3.7); 
 
 \draw[dashed,red](7.5,3,1) -- (7.5,3,3.7); 
 \draw[dashed,red](7.5,1,1) -- (7.5,1,3.7); 
 
 \node[text width=3cm] at (5.4,1,.5) { $i_b$};
 \node[text width=3cm] at (-2.8,.5,-.6) { $j_a$};
 \node[text width=3cm] at (-2.8,2.5,-.6) { $j_b$};
 
 \node[text width=3cm] at (-2.8,.5,2.7) { $k_b$};
 
 \draw[->,blue] (1.5,1,1.5) -- (1.5,2,1.5);
 \draw[->,blue] (3,1,1.5) -- (3,2,1.5);
 \draw[->,blue] (4.5,1,1.5) -- (4.5,2,1.5);
 \draw[->,blue] (6,1,1.5) -- (6,2,1.5);
 
 \draw[->,blue] (1.5,1,2.5) -- (1.5,2,2.5);
 \draw[->,blue] (3,1,2.5) -- (3,2,2.5);
 \draw[->,blue] (4.5,1,2.5) -- (4.5,2,2.5);
 \draw[->,blue] (6,1,2.5) -- (6,2,2.5);
 
  \node[text width=3.3cm] at (-2.8,1.6,5) { Plane wave generated at this wall};
  
  \draw[->,dashed] (-2.8,1.6,4.5) -- (1.5,1,3);






\end{tikzpicture}
    \caption{Total/scattered fields in three dimensions.}
\end{figure}
The total scattered electric field density in Y direction can be obtained at the two ends by
\begin{align}
D_y\left(i,j+\frac{1}{2},k_a\right) = D_y\left(i,j+\frac{1}{2},k_a\right) - 0.5 H_{x,inc}(j),
\end{align}
\begin{align}
D_y\left(i,j+\frac{1}{2},k_b+1\right) =& D_y\left(i,j+\frac{1}{2},k_b+1\right) \\&- 0.5 H_{x,inc}(j),
\end{align}
The simulation of the Gaussian pulse is given by Fig.~21 while the analytic solutions are the same as FDTD-2D.
\section{Conclusion}
In this paper, we present a global framework analysis of FDTD 1D, 2D and 3D considering various combination of the medium properties such as absorbing boundaries, perfect matched layer, dielectric cylinder, etc. We provide pictorial presentations of the wave propagation in various instant to confirm the behavior of the wave in the medium with the analytic solution. In addition, we present the design of the patch antenna and the optimal parameters that provide the required antenna properties. Capitalizing on these parameters, we derive the antenna figure of merits such as the pattern, the directivity at both azimuthal and elevation planes, the input impedance, the VSWR, the return loss, and the $|S_{11}|$ parameter. As a future direction, we intend to design a large array antenna with powerful bandwidth to operate in the 5G Millimeter wave frequency band.


\bibliographystyle{IEEEtran}
\bibliography{full}

\end{document}